\documentclass[12pt]{article}
\usepackage{amsmath}
\usepackage{amsfonts}
\usepackage{epsfig}
\usepackage[english]{babel}
\usepackage[T1]{fontenc}
\usepackage[latin9]{inputenc}
\usepackage{amsthm}
\usepackage{amssymb}
\usepackage{graphicx}
\usepackage{epstopdf}
\usepackage{subfig}
\usepackage{color}
\usepackage{bbm}
\usepackage{mathtools}
\usepackage{hyperref}
\usepackage{natbib}
\usepackage{nicefrac}

\graphicspath{{Figures/}}

\begin{document}

\title{A new Model of City Growth and its Application to a middle sized French City}
\author{Athanasios Batakis \and Nga Nguyen \and Michel Zinsmeister}

\maketitle

\begin{abstract}
This paper consists of two parts: in the first one, after a brief history of city growth modelling, we introduce a new theoretical model based on correlated percolation. In the second part we apply this model to a concrete town, the city of Montargis, for which we could dispose of all the urban history since 1900. 
 It appears that the embedded algorithm is quite efficient in terms of computational time and allows to exploit large data type ressources such as for instance individual land lots.
\end{abstract}


\section{Introduction}
In 2022, more than half of the world population lives in cities. This is why
modeling urban growth has become an important challenge. The aim of this work is  to better understand and eventually predict urban development in order to supply tools for decision makers. 

In a celebrated paper of \cite{harris}, the city is  considered as a multi-centered system of clusters with different hierarchical levels. In fact, for all cities, the centers, suburbs and hinterlands which form this hierarchy share the same features, both in terms of functionality and topographic patterns. Their functional scheme is repeated across different spatial scales, in a self-similar pattern: districts of different sizes at different levels in the hierarchy have a similar structure. 
Moreover, city growth not only occurs through the addition of units of development at the most basic scale, but also through increasing specialization of key centers. This analysis strongly suggests that urban growth is in some sense an intelligible process, hence accessible to various quantifications.\\

Size of cities  depends on  local interaction of populations, production of goods, commercial activities, and various other factors. Nevertheless, even the largest cities grow from the tiniest seeds which, repeated at different levels of hierarchy, interact in different scales in a self-similar way \citep{P.Frankhauser}, making pertinent the use of fractal geometry tools.\\

In the middle of the 1980's, \cite{M.Batty86} have been the first to apply fractal geometry to study and model city growth. A few years later, \cite{P.Frankhauser} proposed a more general overview of this geometry in urban geography. Fractals are used to define urbanisation rules applied at different scales of the urban areas (e.g.  \cite{frankhauser2008,frankhauser2010,frankhauser2011,badariotti2005}).\\

The urban structure not only undergoes fractal properties but also, like all natural objects, grows and evolves with time. Any model of city growth must thus involve a dynamical aspect.\\
In fact, cities evolve through the cumulative addition and deletion of basic units, cells or particles on a determined space. Such units may be buildings, population, transportation networks,... All of them co-exist, interact in a urban system, and embody its dynamical growth.\\

A first attempt to describe urban expansion through mathematical models of fractal growth goes back to \cite{M.Batty}. They adapted a well known model, called  diffusion limited aggregation (DLA). This model, proposed by Witten and Sander in 1981, is a process whereby particles undergoing a random walk cluster together to form aggregates. It is a simple model that was able to generate fractal structure whose self-similarity was dendritic or tree-like. In \citep{M.Batty}, the authors have calculated the fractal dimension of some urban structures and noticed that most of them are  lying between $1.6$ and $1.8$, around the empirical value of dimension of DLA cluster, namely $1.71...$. Even if this numerical evidence pleads in favor of their approach, this model has to be rejected since it does not fit with some other observed properties of urban systems. Most of urban structures are actually more compact near their center than DLA clusters. 
Furthermore no-interaction between clusters occurs while a city is actually a complex system of related components.\\
 
In order to overcome some of the limitations of DLA model, \cite{Batty} suggested to use the dielectric breakdown model (DBM), a modified version of DLA, in an attempt to model the urban growth of Cardiff. The DBM model, developed by Niemeyer, Pietronero, and Weismann in 1984, is a combination of DLA and previously existing electric field  models. It allows to generate cities of different shapes and degrees of compactness. 

Despite the fact that  simulations of DBM are more realistic, they remain limited. A cluster generated by this model does not satisfy the complex property of an urban system: its main drawback concerns its  density distribution. The mean density of a DBM-cluster in simulations decreases from the center as a power law 
\[
\rho(r)\sim r^{D-2}
\]
where $r$ is the radial distance from center of cluster, and $1<D<2$ is the fractal dimension of the simulated cluster. 

But, even if this is disputed in particular for modern megacities, it is commonly admitted that most  cities obey Clark's law (\citep{C.Clark}) saying that $\rho(r)$, the mean density of a city at distance $r$ of its center decays exponentially, more precisely
\[
\rho(r)\sim e^{-\lambda r}
\]
where $\lambda$ is the so called \textit{density gradient}.\\
A word has to be added about the definition of density: Clark means density of population and the definition of \textit{local} density has to be precised as we will. Also the density of population maybe very different from the density of landlots, especially in big cities with high buildings. However, since our main object of study is a quite small town for which land lots are roughly equally populated, we will consider in our models density as density of land lots. Nevertheless our models maybe easily adapted to situations where the two notions of density differ.

Before we come to the main approach of this paper let us quote some other models of city growth.
One of the main streams of modeling is Cellular Automata introduced by Stanislav Ulam and John von Neumann in 1943. \cite{Tobler} was the first to apply CA in this context, followed by \cite{Sante}, \cite{Li}. For more references see \cite{Battymultiple}.

Other models of city growth are of more econometric nature, such as the Random Utility Models introduced by \cite{McFadden} in 1974.

Our approach is a continuation  of \cite{H.Makse1998} in which a toy model based on correlated gradient percolation is proposed to simulate Berlin with remarkable results. The model is time evolving and generates more complex urban areas which turn to be more realistic than the preceding ones. Compared with "real" Berlin, it gives of course another "random" Berlin, but the impression of an air of family is significant, suggesting that some cities could grow according to some stochastic law.

However, this toy model faces two drawbacks. First, it doesn't take in account all possible kinds of city growth. Secondly, it is too rough to describe a concrete situation. For instance, in our example, the city of Montargis clearly does not grow as Berlin.

In the model of \cite{H.Makse1998}, distance to the center governs city  growth  and dynamics of the model is guided by the density gradient $\lambda$  playing the role of the inverse of time. 

%

The main goal of this paper is to describe a new theoretical model still based on correlated percolation but for which the role played by the distance to the center in \cite{H.Makse1998} is overtaken by local density of buildings at current time. In a second step, we push further this idea, replacing local density by a quantity leading to the local density in expectation at next step.
This model may be run forward or backward, making possible to check the validity of its predictions. \\

In the second part of the paper we propose a concrete application of this new model to the city of Montargis.  We adapt it by taking into account other factors that affect city growth, the most important among them being the concept of accessibility. We claim that this new parameter is better suited to towns with high dispersion which is the case of Montargis.\\
The simulations are carried out on land lots; we have compared simulations with real city data and obtained fairly good results.\\

The paper is organized as follows:\\
In the second section, we present some mathematical background about (un)-correlated gradient percolation and describe our new theoretical model of city growth . \\
The third section is devoted to the case-study of Montargis, a middle-sized French city for which, after a quick geographical description, we describe two new concrete models together with the results of actual simulations.

\section{Percolation Models}
Percolation is a random model that was introduced by John Hammerslay in 1957 in order to describe the flow of a fluid or a gas through a porous medium with small channels which may or may not let fluid  pass. During the last decades, percolation theory has brought new understanding and techniques to a broad range of topics in physics, materials science, complex networks, geography,.... 
Percolation theory describes the behavior of randomly formed clusters in a grid. Each site of the grid is colored black with probability $p$ or else white, the different sites being independently colored. We call clusters the connected components of the set of black sites.\\
There exists a well defined \textit{critical value} $p_c$ such that:
\begin{itemize}
\item if $p<p_c$ there is no infinite cluster a.s., this is called \textit{sub-critical} regime; 
\item if $p>p_c$ there is a unique infinite cluster \textit{super-critical} regime a.s.
\end{itemize}
\cite{H.Kesten} proved that in 2-dimensions, on triangular lattice, $p_c = 1/2$.\\

The value of $p$ being given, this model provides a set of occupied (black) sites: in order to make it dynamical one lets $p$ increase to $1$. As $p$ crosses the critical regime $p=p_c$, the unique infinite cluster starts to look like a "city" if we think of occupied sites as buildings. But this model is irrealistic: these "cities" have no center and no border!\\
To make up for this drawback we use an alternative notion of percolation, namely gradient percolation.

\subsection{Gradient Percolation}
\textit{Gradient Percolation} is a model of inhomogeneous percolation introduced by \cite{B.Sapoval}. In this model, the occupation probability $p$ attached at each site $z$ is a function of $z$ decreasing from $1$ to $0$ as the distance from $z$ to the origin increases from $0$ to $\infty$.

In what follows we will systematically adopt the complex notation $z$ to denote a site (pixel) in the ``landscape''. 

The existence, size and position of a cluster of occupied sites in the region depend on the set of $z$ such that $p(z)$ is greater than the critical value $p_c$. The cluster containing the origin (=the center of the city) has a border close to the set of points where $p(z)$ becomes less than the critical value $p_c$. 

As already mentioned, according to data of numerous cities, \cite{C.Clark} claims that the population density $\rho(z)$ of the real urban systems should satisfy the relation
\begin{equation}
\label{eq:density}
\rho(z) = \rho_0e^{-\lambda r_z},
\end{equation}
with $r_z = d(z,0)$ is the distance from $z$ to the center $(=0)$ of the lattice-city, where $\lambda$ is the so called \textit{density gradient}.

Therefore, we assume that the occupation probability is an exponential function of $r_z$:
\begin{equation}
p(z) = \frac{\rho(z)}{\rho_0}=e^{-\lambda r_z}.
\label{eq:2prob}
\end{equation}
The density gradient $\lambda$ influences the size of the central cluster. With small $\lambda$ ($\lambda\approx 0$), the central cluster is large (corresponding to  developed cities), and vice versa, the larger the value of $\lambda$ ($\lambda \approx 1$) is, the smaller the central cluster is. As before, we get in this way a growth process where the role of time is played by ${1}/{\lambda}$.\\
This modification of the percolation process takes into account Clark's law but does not rule out the fact that "cities" obtained in this way have a rotational symmetry for example, which is rarely the case of concrete cities. Neither does this model take care of Christaller law saying that growing cities generate "satellite" smaller centers, a well-observed fact. In order to take into account these last points we need another modification explained in the next paragraph.  

\subsection{Correlated Gradient Percolation} 

Observation of real cities suggests that new constructions are not positioned independently of the existing ones. The positions/parcels  seem to be occupied with probability depending on the presence of  neighbors. In other words, there exists a correlation between occupied land lots: near an occupied site, the probability of additional development is higher and decreases as we move away from it. We argue that these assumptions of site occupation affect growth of urban areas and thus introduce the following \textit{correlated gradient percolation} model.\\

Such a model is based on a discrete Gaussian field, i.e. an bi-infinite (un)correlated Gaussian matrix. The uncorrelated case corresponds to the \cite{B.Sapoval} model.

In order to build a bi-infinite matrix of correlated random numbers depending on the distance between two sites,  we start with $(X_z),\,z\in \mathbb{Z}^2$  an infinite matrix of independent variables (i.e. uncorrelated) with identical distribution $\mathcal{N}(0,1)$, the standard normalized Gaussian distribution. From this Gaussian matrix we wish to construct another one, say $(Y_z)$ with covariant matrix given by
$$\mathrm{cov}\left( Y_z,Y_z'\right)=C(\vert z-z'\vert),\;z,z'\in\mathbb{Z}^2,$$
where 
$$C(\ell)=\frac{1}{(1+ \ell^2)^{\alpha/2}}.$$
The parameter $\alpha$ has to do with the concentration, in the sense that the smaller the value of $\alpha$ is, the higher will be the concentration of clusters.\\
We look for such a matrix in the form
$$Y_z=\sum_{z'\in \mathbb{Z}^2} a(z'-z)X_{z'},$$
for which we may compute, assuming $a$ is even,
$$a*a(z)=C(\vert z\vert),$$
where $*$ stands for convolution of functions defined on the group $\mathbb{Z}^2$. This equation may be solved by taking Fourier transform of both sides,
$$\hat{a}(t,t')^2=\hat{S}(t,t'),$$
where $S(z)=C(\vert z\vert)$, an equation which has a solution since $\hat{C}>0$ (actually $\hat{C}$ has been exactly computed in \cite{H.Makse1996})\\
To finish the construction, we denote first by $F$ the cumulative distribution function of a random variable following a $\mathcal{N}(0,1)$-law: the numbers $\eta(z)=F(Y_z)$ are then random numbers uniformly distributed in $[0,1]$, and they form the desired matrix.

\subsubsection{Makse et al's Model}
\noindent
For a site $z$ on a square lattice of size $M\times N$, we consider the following occupation probability function:
\begin{equation}
p(z) = e^{-\lambda r_z}
\label{eq:pro1}
\end{equation}
where $\lambda$ is is the density gradient and $r_z = |z|$ is the distance from $z$ to center $0$ of the  lattice.\\
We generate on the lattice a matrix $(\eta(z))$  from a correlated Gaussian matrix as in the preceding paragraph.\\
\noindent
On the square lattice $L$ of size $M\times N$, a site $z$ is black and marked $1$ (occupied) or white with marked $0$ (vacant) with probability given by \eqref{eq:pro1} in the  following way:
\begin{equation}
L(z) = \begin{dcases*}
        1  & if $\eta(z)<p(z)$\\
        0 & otherwise,
        \end{dcases*}
  \label{eq:matric}
\end{equation}
\noindent
This model has been developped by \citep{H.Makse1996} and succesfully used to model the city growth of Berlin. This success is partly due to features that are particular to Berlin, which is a city with a big center with high attractivity (so that the distance to the center plays a proeminent role) and rather rotationally symmetric, with no big river crossing it, for instance.\\
Thus, in order to deal with other kind of cities, we need to further modify the model. 

To summarize,  a correlated gradient percolation model  (CGP) consists of 2 objects :
\begin{enumerate}
\item[1)] a Gaussian correlated field, with correlation parameter $\alpha$ as defined above,
\item[2)] a threshold function depending on the site $z$ and on time $t$.
\end{enumerate}

In the case of  \citep{H.Makse1996} the threshold function $p_t(z)=e^{-\frac{c}{t}r_z}$ depends only on the ratio distance to the center (of the city) over time.

\section{New CGP Models}
In this section we will describe CGP models based on  the same Gaussian field but with different and more complex threshold functions.

 In particular, $p_t(z)$ will depend on the realization of the city cluster at time $t$. 
 
The threshold function in this new models becomes thus a ${\mathcal F}_t$-measurable random variable, where ${\mathcal F}_t$ is the $\sigma$-algebra of events up to time $t$.
 
\subsection{First modified CGP-model}\label{firstMCGPM}
Let us first introduce some notations and concepts for this new model. A town will be a finite subset of $\mathcal{S}:=\{1,...,M\}\times\{1,...,N\}\subset \mathbb{Z}^2$. The set $\mathcal{M}\subset\mathcal{S}$ consists of the set of black points (corresponding to the buildings). Equivalently it is a binary $M\times N$ matrix $\left({\mathcal M}(z)\right)$. The growth model that we are  going to consider consists in iterating an algorithm that we now describe.

\noindent
We call $\mathcal{M}_t$ the town at time $t$ and wish to construct $\mathcal{M}_{t+1}\supset\mathcal{M}_t$. We consider a function $p_t: \mathcal{S}\setminus \mathcal{M}_t\rightarrow [0,1]$ and $\mathcal{G}$ a correlated Gaussian $M\times N$ matrix whose entries follow a $\mathcal{N}(0,1)$ law.

\noindent
Let $\eta=\mathcal{G}(\omega)$ be a realization of the random matrix. For all $z\in\mathcal{S}\setminus\mathcal{M}_t$, we decide to include $z$ in $\mathcal{M}_{t+1}$ if and only if $\eta(z)<p_t(z)$ as in the above described model of correlated gradient percolation. We then consider a function $p_{t+1}:\mathcal{S}\setminus\mathcal{M}_{t+1}\rightarrow [0,1]$ such that $p_{t+1}> p_t$ and we start over the process. The novelty in this model is that we allow $p_{t+1}$ to depend on $\mathcal{M}_t$.

\noindent
This basic structure being described, the challenge is now to choose the function $p_t$, as well as the parameter $\alpha$ and to adjust them in order to fit with reality.

\noindent
Concerning the function $p_t$ we already mentionned the choice made by \cite{H.Makse1998}: $p_t=e^{-c\frac{r_z}{t}}$ where $t$ is now a continuous parameter going from $0$ to $\infty$ and $r_z$ is the distance of $z$ to the center of the city. This choice induces very "compact" towns for which the only criterium of choice is the vicinity of the center.

\noindent
Partly because of our main subject of study, namely the city of Montargis (see below), we prefer to privilege more decentralized towns by taking $p_t$ to be the concentration or density function. This is defined as
\begin{equation}\label{pi}
p_t(z)=1_{\mathcal{M}_t}\ast \varphi(z)
\end{equation}
where $\varphi$ is a nonnegative function with 
\begin{equation*}
\sum_{z\in\mathcal{S}}\varphi(z)=1,
\end{equation*}
and the convolution ($\ast$) is the (already defined) one on $\mathbb{Z}^2$ after having extended the two functions by $0$ outside $\mathcal{S}$. In order to fit with Clark's law we only consider functions $\varphi$ of the form $\frac{1}{c'}e^ {-c|z|}$.

\begin{figure}[h!]
	\centering
	  \subfloat[Baltimore 1925]{\label{fig:bal1925}\includegraphics[width=0.5\textwidth]{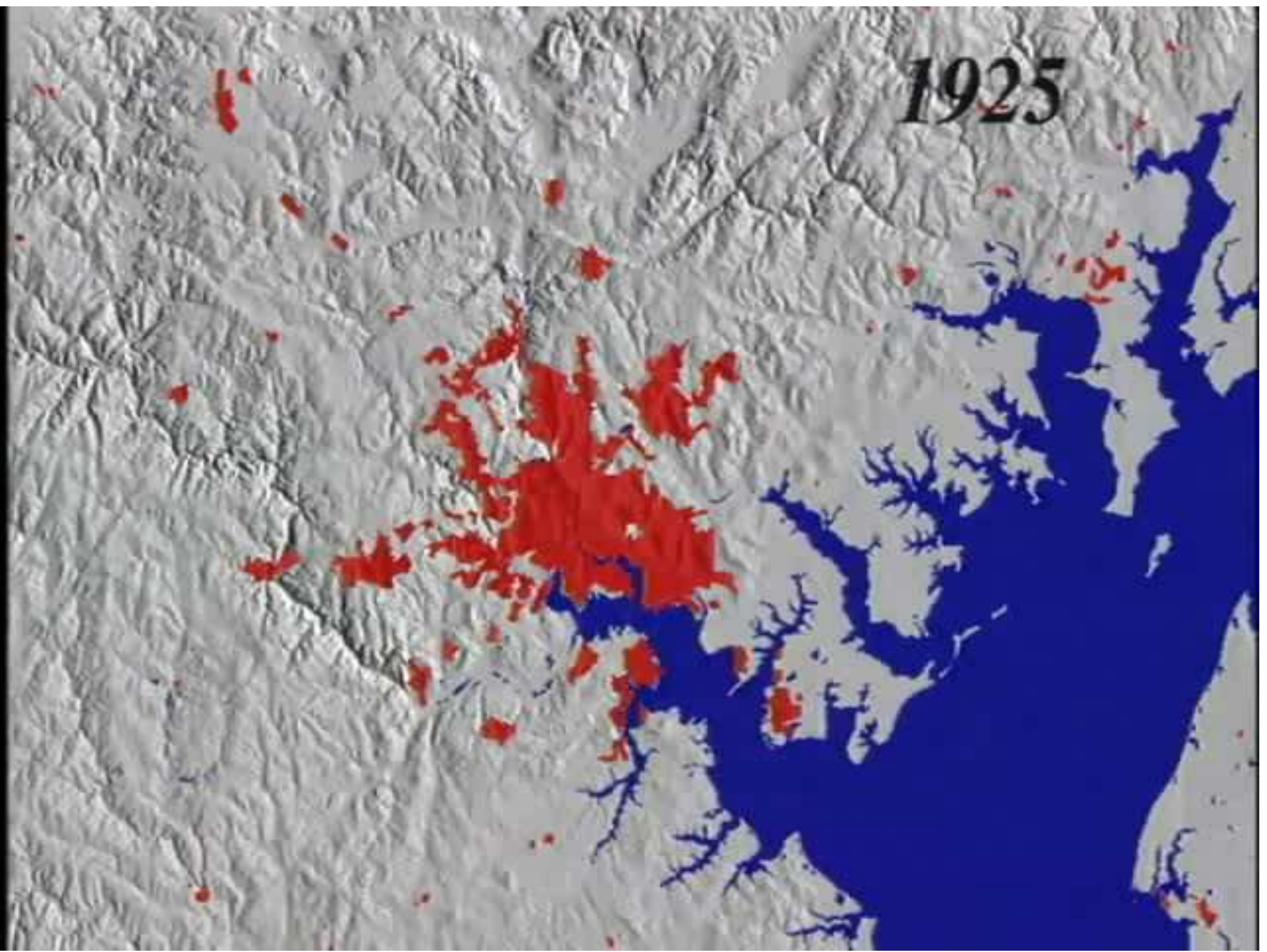}}                
  \subfloat[Simulation ]{\label{fig:simbal5}\includegraphics[width=0.5\textwidth]{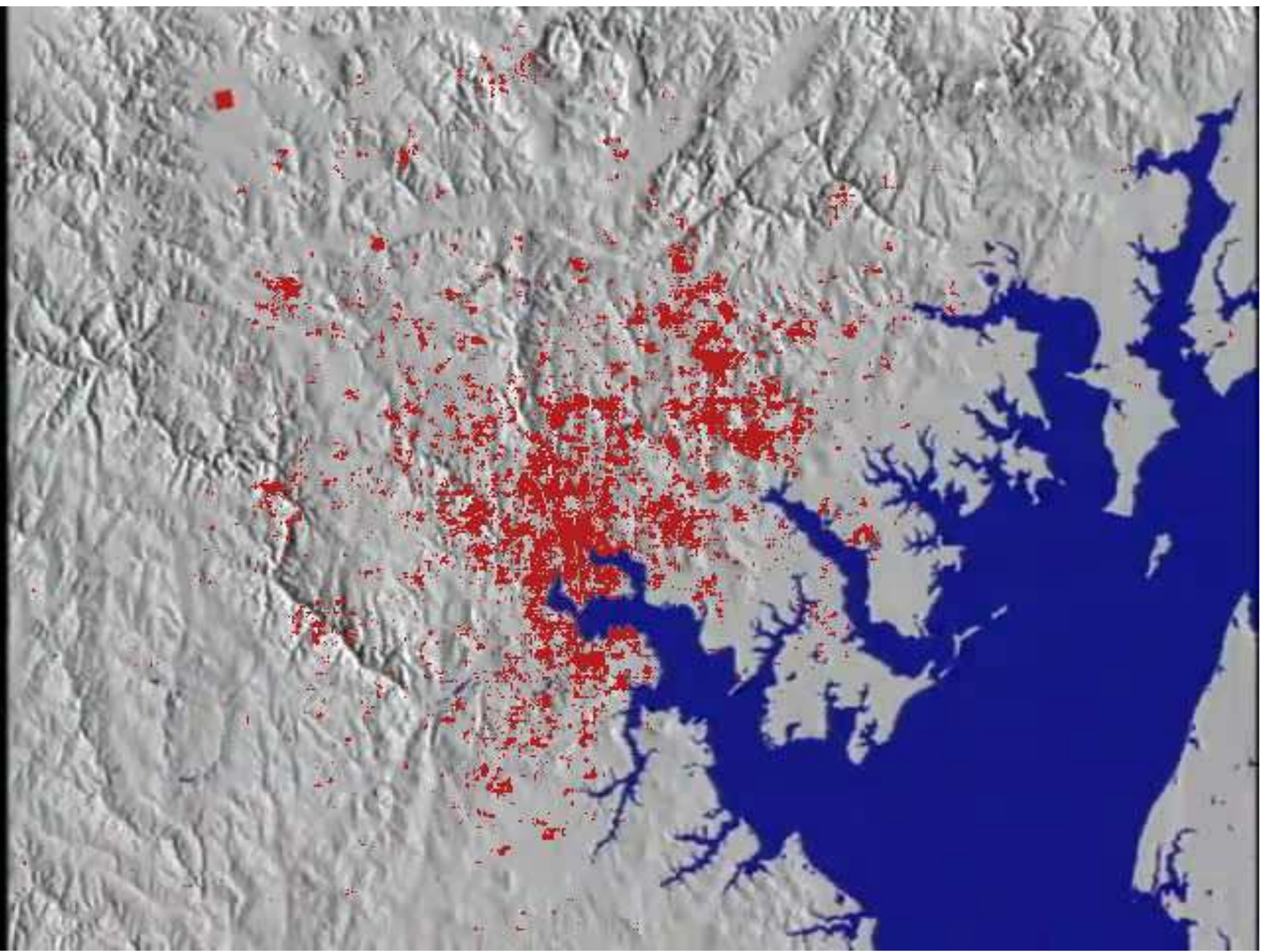}}  \\
	  \subfloat[Baltimore 1992]{\label{fig:bal1992}\includegraphics[width=0.5\textwidth]{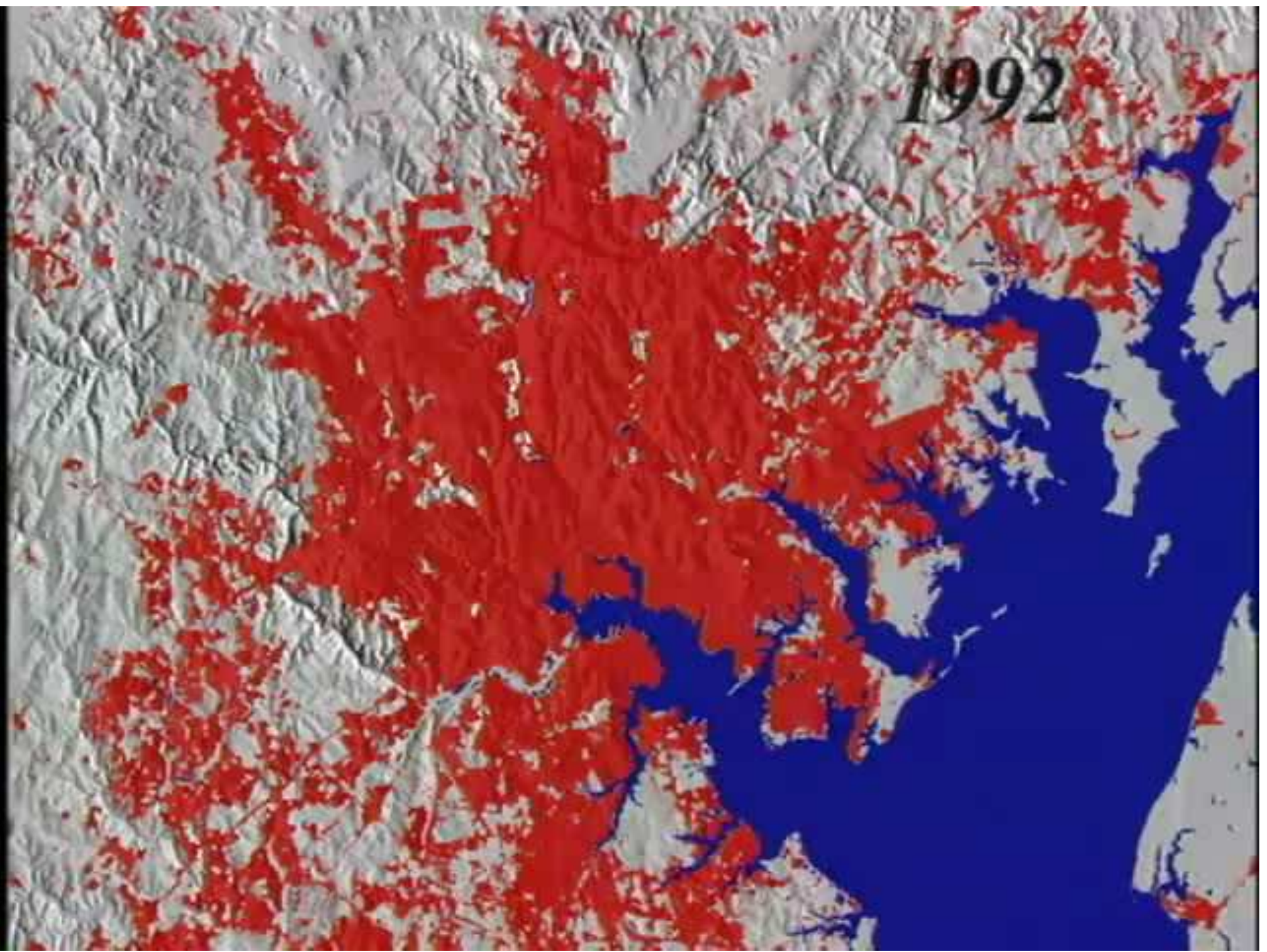}}     
		\subfloat[Simulation ]{\label{fig:simbal11}\includegraphics[width=0.5\textwidth]{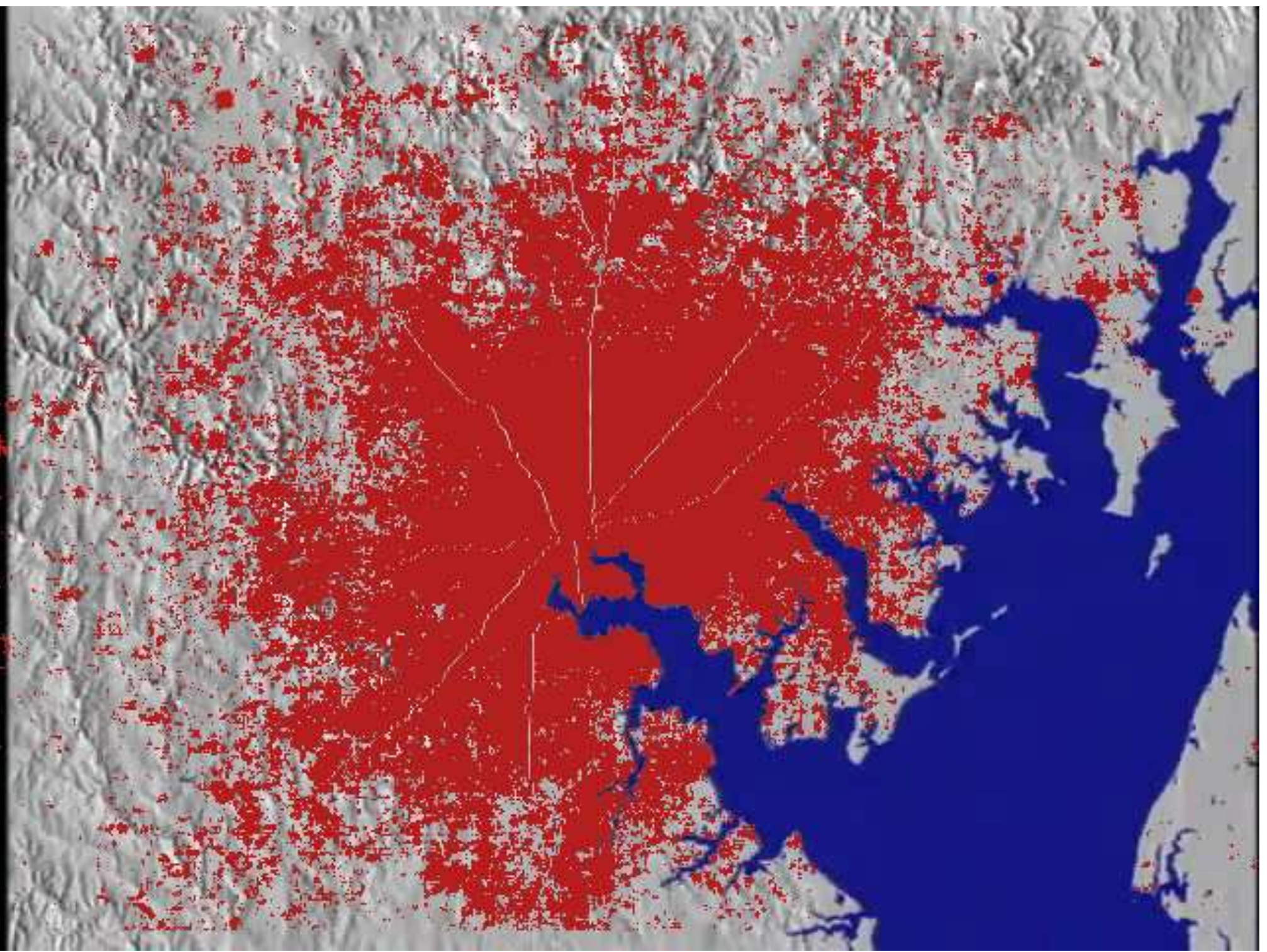}}
  \caption{Real Baltimore and its simulation}
  \label{fig:resultBal4}
\end{figure}

Running this toy-model for the city of Baltimore (for which all data is available on the net) starting from the actual city as early as in 1792 we have obtained the following simulations which we compare with reality, fig. \ref{fig:resultBal4}. 

It would have been of course very surprising to get the same picture but, as in \cite{H.Makse1998} and the case of Berlin, the simulations make plausible that Berlin and Baltimore's growth obey some probability law in the same sense for instance that two realizations of DLA clusters look alike without being the same.

This leads us to make the hypothesis that the correlation exponent $\alpha$ might be pertinent to distinguish between different types of city growth. In particular it seems  interesting to estimate statistically this exponent for a given city. This is part of an ongoing work.


%
%

. 

\subsection{Second modified CGP-model}
A slightly different model is based on the belief in harmonious growth of cities. It makes sense for a city which is already well developed and for which we want to predict the future growth, assuming its harmonious character. So let $\mathcal{M}_0$ be the present stage of the city and, as before, $p_0=1_{\mathcal{M}_0}*\varphi$, that is the density function at $t=0$.

\noindent
The idea is not to take $p_0$ as the percolation threshold but instead to take a function $\rho$ to be defined in such a way that, denoting by $\mathcal{M}_1$ the new city built using CGP method with this threshold, we have
$$\rho=\mathbb{E}(\mathcal{M}_1)*\varphi$$
outside $\mathcal{M}_0$. In other words, we want $\rho$ to be equal to the density function of the city it builds, in expectation. This can be seen as a kind of martingale property.\\
In order to compute this function we write $\mathcal{M}_1$ as the disjoint union of $\mathcal{M}_0$ and $\mathcal{H}_1$. We first notice that 
$$\mathbb{E}(1_{\mathcal{H}_1})=(1-1_{\mathcal{M}_0})\rho,$$
so that the equation for $\rho$ is
$$ \rho=p_0+\mathcal{T}(\rho),$$
where 
$$\mathcal{T}(\rho)=\left[ (1-1_{\mathcal{M}_0})\rho\right]*\varphi.$$
Unfortunately (but this is not much of a surprise), the only solution of this equation is the constant $1$. We pass this difficulty by posing
$$\rho_t=(\mathcal{I}-t\mathcal{T})^{-1}(p_0)$$
which defines a growing one-parameter family of thresholds, and thus a growing city as time grows from $0$ to $1$. This model will be used later for the growth of Montargis: we call it the \textit{forward model}.\\

This model may also be run backwards. In order to recover a town in the past we change the above model by using the dual $\mathcal{M}' = \mathcal{S}\setminus\mathcal{M}$, at each site $z$, if $1_{\mathcal{M}}(z) = 1$ then $1_{\mathcal{M}'}(z) = 0$ and vice-versa. The growth model is generated by using this matrix similarly to the forward model. But in this case the white area is growing, so the black one is decreasing, as it should be.

This will be used for the city of Montargis to check the validity of the model and to adjust the parameters.

\section{Application to a concrete case}

In this part we focus on a particular case. Answering a call for tender of the ``R\'egion Centre'' to simulate the expansion of the city of Montargis we wanted to adapt the aforementioned models.
But, to this aim, we needed a much more precise approach made possible by the availability of land lots data, including their history since 1900. This information has also allowed us to integrate a new parameter, namely accessibility, into the model. 

Before we come to the details, we start with a small geographical presentation of the city.

\subsection{A Brief Geographical Presentation of Montargis}
The city of Montargis is the second (after Orl\'eans) most populated town in the d\'epartement du Loiret, France. The whole basin of population reaches 70000 inhabitants, which makes it, in French standards, a middle size town. It is situated 130 km south east of Paris, a proximity that, as we shall see later, plays a role in the economy of the town. Montargis is also 70 km north-east from Orl\'eans, the local economical center. The town is crossed by the river Loing, which is an affluent of the Loire river.\\
In the sixties, due to the proximity of Paris, many people from the Paris area (Ile de France), had built summer houses in the Montargis area. These summer houses became principal ones as their owners retired. This fact participates to the fact that the population is rather dispersed in the vicinity of Montargis, and also rather old. The population of the center is declining, but the population of the basin is slightly increasing, a phenomenon which is shared by most middle-sized french towns.\\
The main local economical activity is low-tech industry (rubber and paper industry), a fact that may explain that as far as median income is concerned, Montargis is one of the less favoured cities of this size in France. In order to illustrate the long-run industrial tradition of Montargis, let us mention that in the 20's, in the framework of a program called \textit{Work and Study}, several future leaders of the Chinese Communist Party have stayed and worked there. The most famous one is Deng Xiaoping who has spent two years (1923-24) as a worker in the Hutchinson factory which still exists. For the record this fact has generated a Chinese touristic movement with people coming every year to visit the places (that have been carefully preserved) where Deng lived and worked.\\ 
 The proximity of Paris has a second effect: more and more people, while living in Montargis, work in the Paris area, and an even more recent phenomenon (after the pandemy) is remote work, the proximity of Paris facilitating a one or two days travel to the employer in Paris.\\
This little presentation explains in particular why the growth of Montargis is very different from the Berlin one, and our simulations had to take this difference into account.\\

\subsection{Simulations using the forward model}
\begin{figure}[h!]
  \centering
  \subfloat[The rivers of Montargis]{\label{fig:riversMon}\includegraphics[width=0.3\textwidth]{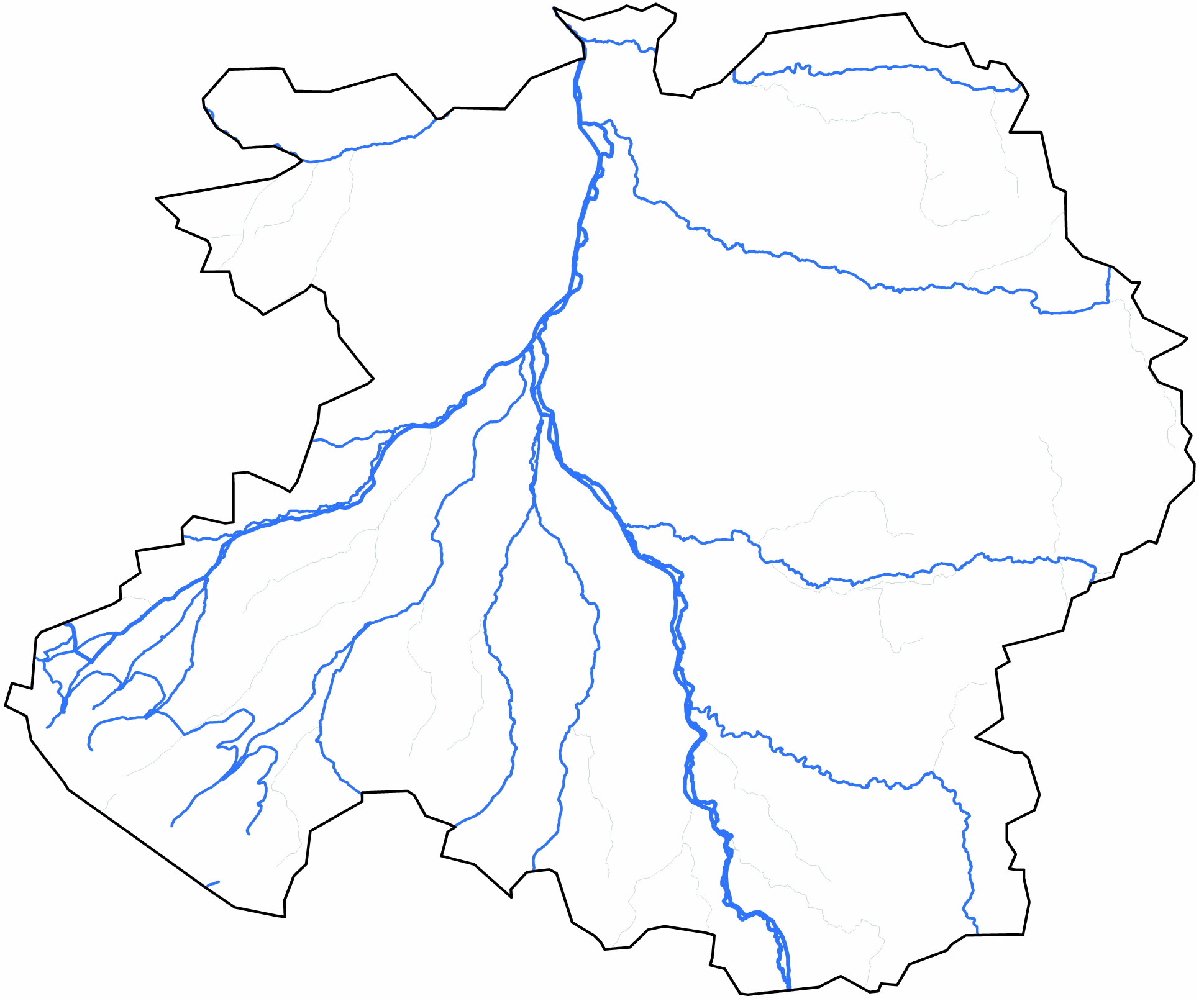}}
   \subfloat[Montargis in 1900]{\label{fig:Mon1900}\includegraphics[width=.3\textwidth]{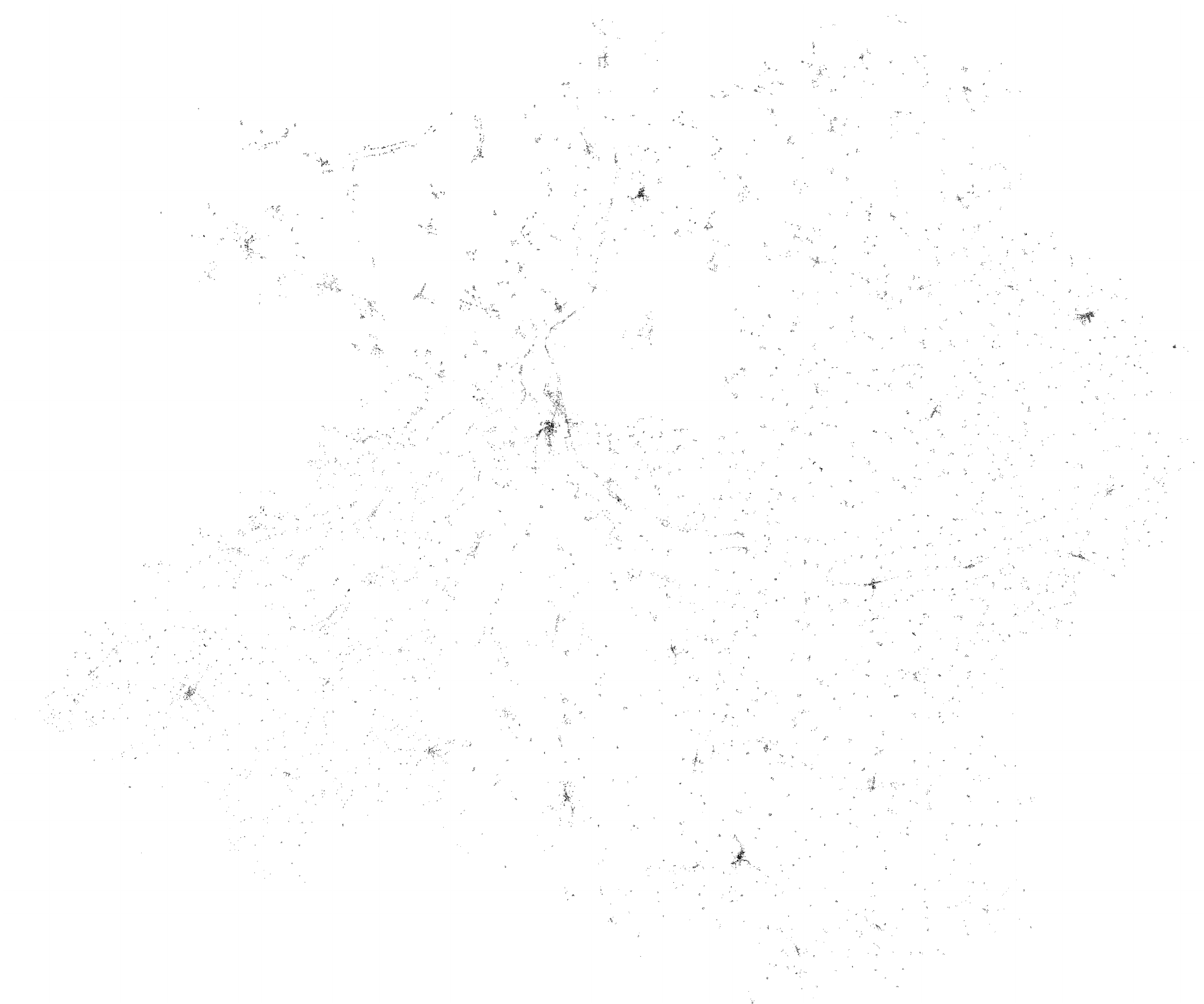}}
  \subfloat[Montargis in 2007]{\label{fig:Mon2007}\includegraphics[width=0.3\textwidth]{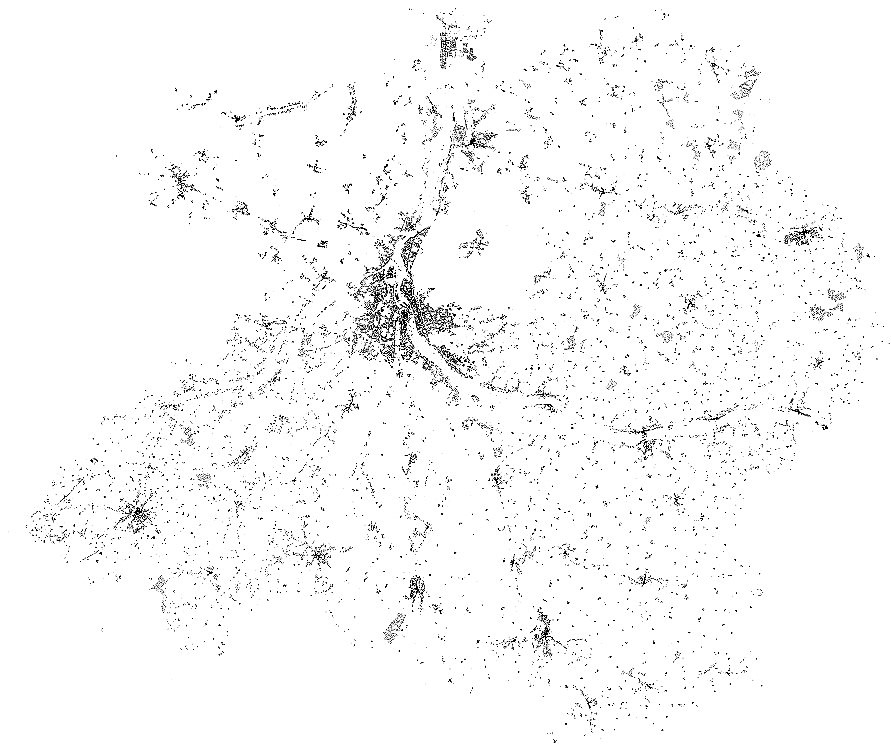}}                    
  \caption{The river and the constructed buildings of Montargis at 1900 and 2007, the data are taken from the project TRUC.}
  \label{fig:dataMol1}
\end{figure}

These simulations are based on the land register of Montargis to which we obtained access. The area of study has been pixelized. A pixel is black if it is in touch with a building. We assume that the role played by the rivers (figure~\ref{fig:riversMon}) in the model is the same as for the buildings. 

The  correlation factor has been empirically fixed to be $\alpha=0.001$, a value that seems to fit the historical growth.

\begin{figure}[h!]
  \centering
  \subfloat[Montargis in 2007]{\label{fig:simforw1}\includegraphics[width=0.5\textwidth]{I0rnew}}                
  \subfloat[$t=0.01$]{\label{fig:simforw2}\includegraphics[width=0.5\textwidth]{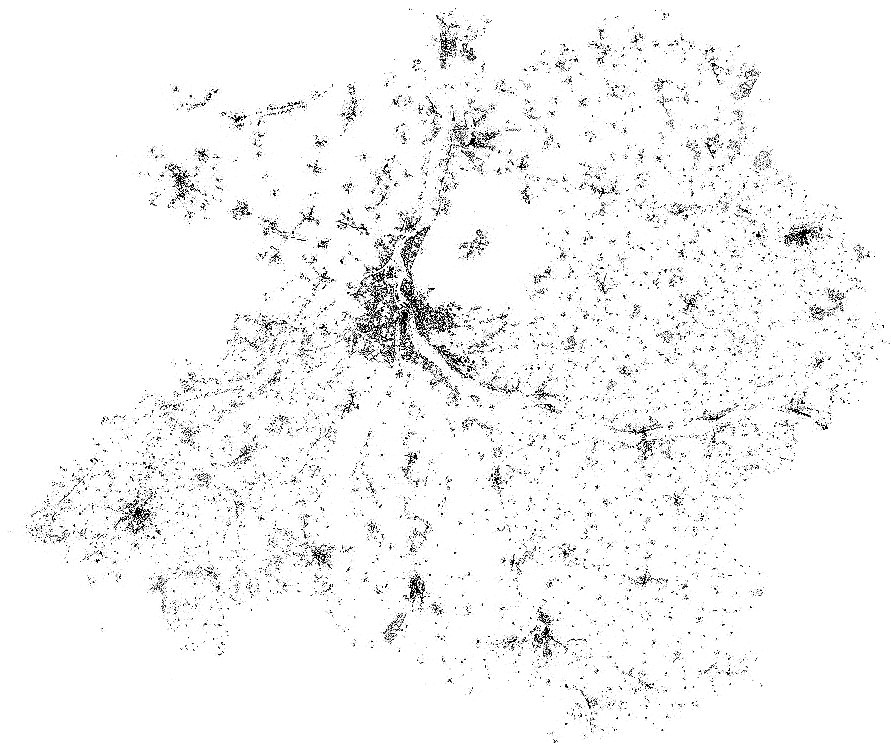}}\\
  \subfloat[$t=0.02$]{\label{fig:simforw3}\includegraphics[width=0.5\textwidth]{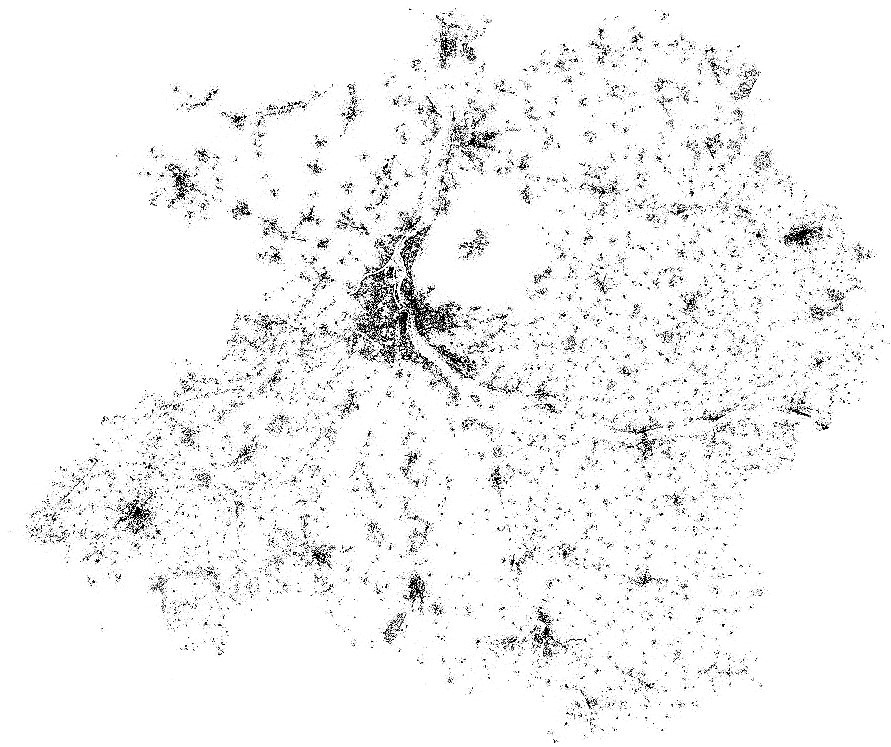}}                
  \subfloat[$t=0.03$]{\label{fig:simforw4}\includegraphics[width=0.5\textwidth]{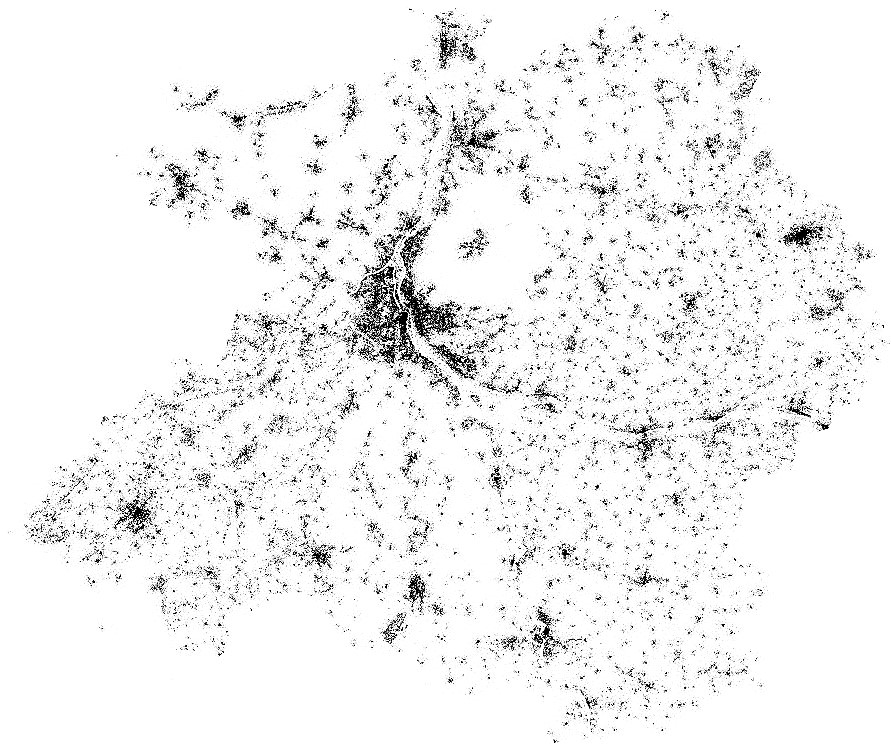}}     
  \caption{The simulations in future of Montargis using forward model in different ``times'' $t$.}
  \label{fig:resultMonForw}
\end{figure}

\noindent
The starting point $t=0$ of the simulations is  Montargis 2007 (figure~\ref{fig:Mon2007}). The simulations are done with a time sequence of scale $\delta t=0.01$, we get figure~\ref{fig:resultMonForw} for $t=0.01,\,0.02,\,0.03$.

\noindent

\begin{figure}[h]
  \centering             
  \subfloat[Montargis in 2007]{\label{fig:simforw0}\includegraphics[width=0.5\textwidth]{I0rnew}}                
  \subfloat[``backward-forward'']{\includegraphics[width=0.5\textwidth]{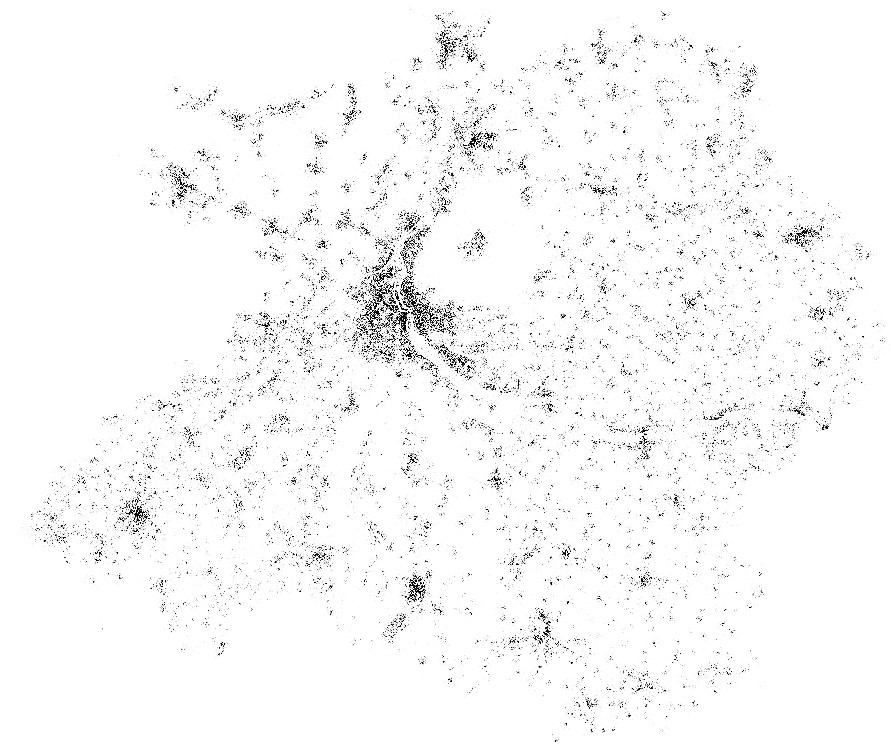}}  
   \caption{Checking the validity}
  \label{fig:resultMonForwB}
  \end{figure}

Since we do not have data posterior to 2008, in order to check validity of the model we perform a backward/forward operation. Starting from Montargis 2007 ($t=0$) we consider t=-0.02 followed by t=0.02. The figure \ref{fig:resultMonForwB} illustrates the result which seems rather convincing.

We have also run the model starting with Montargis 1900 to simulate 2007. The result is essentially the same.

%

\section{A multi-parameter model}
In the preceding model, we assumed that the distribution of buildings models the growth in the future or the reduction in the past on Montargis. In reality, there are many factors interacting in the urban system such as the minimum area around a building, the accessibility, the commercial areas, the schools, the forests, the risk areas .... We will try to include these factors in a modified version of our first model (first modified CGP model, \S \ref{firstMCGPM}) in order to better simulate the growth of Montargis.

\subsection{Model}
\begin{figure}[h!]
  \centering
  \subfloat[Land lots]{\label{fig:parcelles}\includegraphics[width=0.33\textwidth]{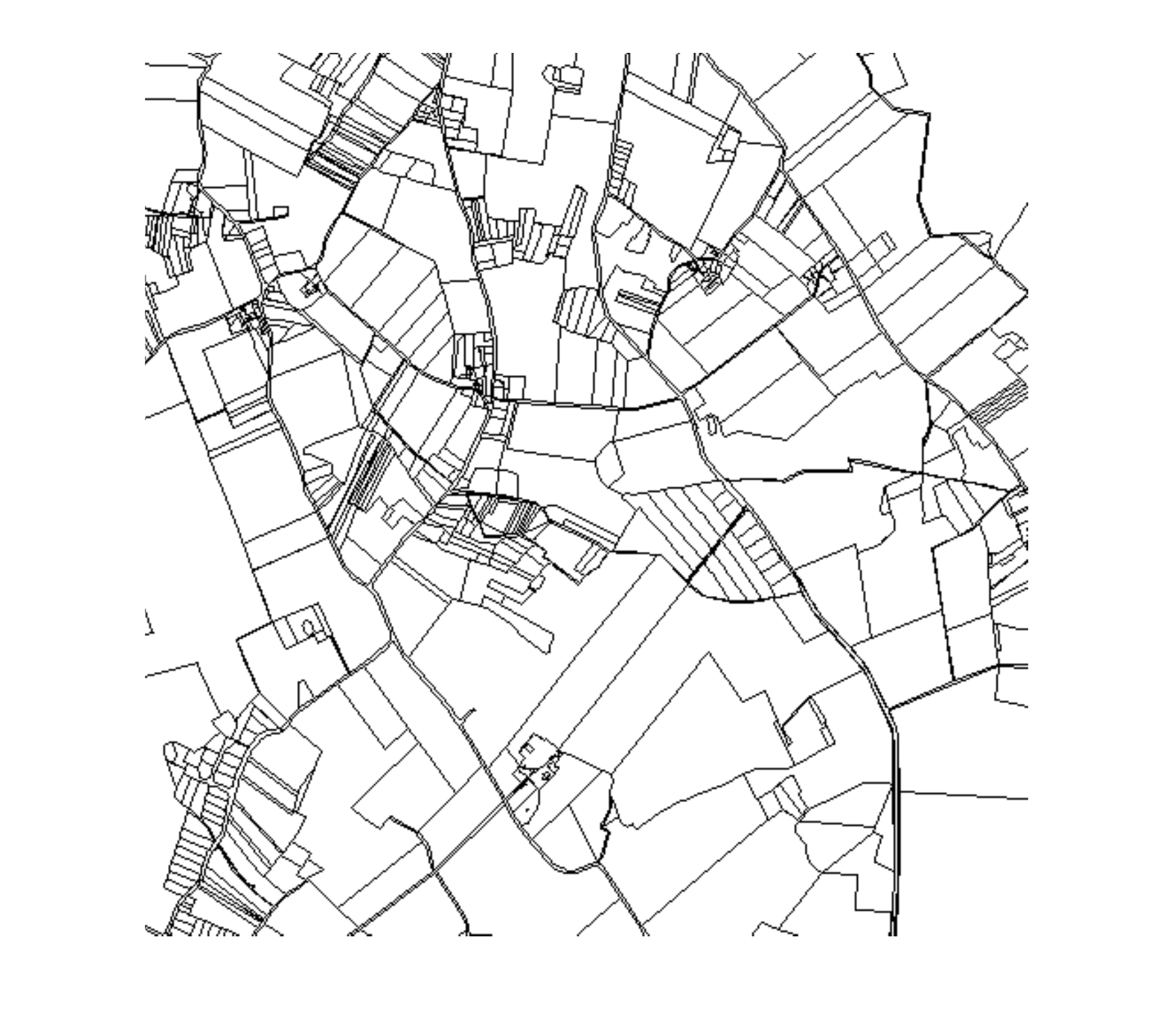}}
  \subfloat[Accessibility]{\label{fig:noeuds}\includegraphics[width=0.33\textwidth]{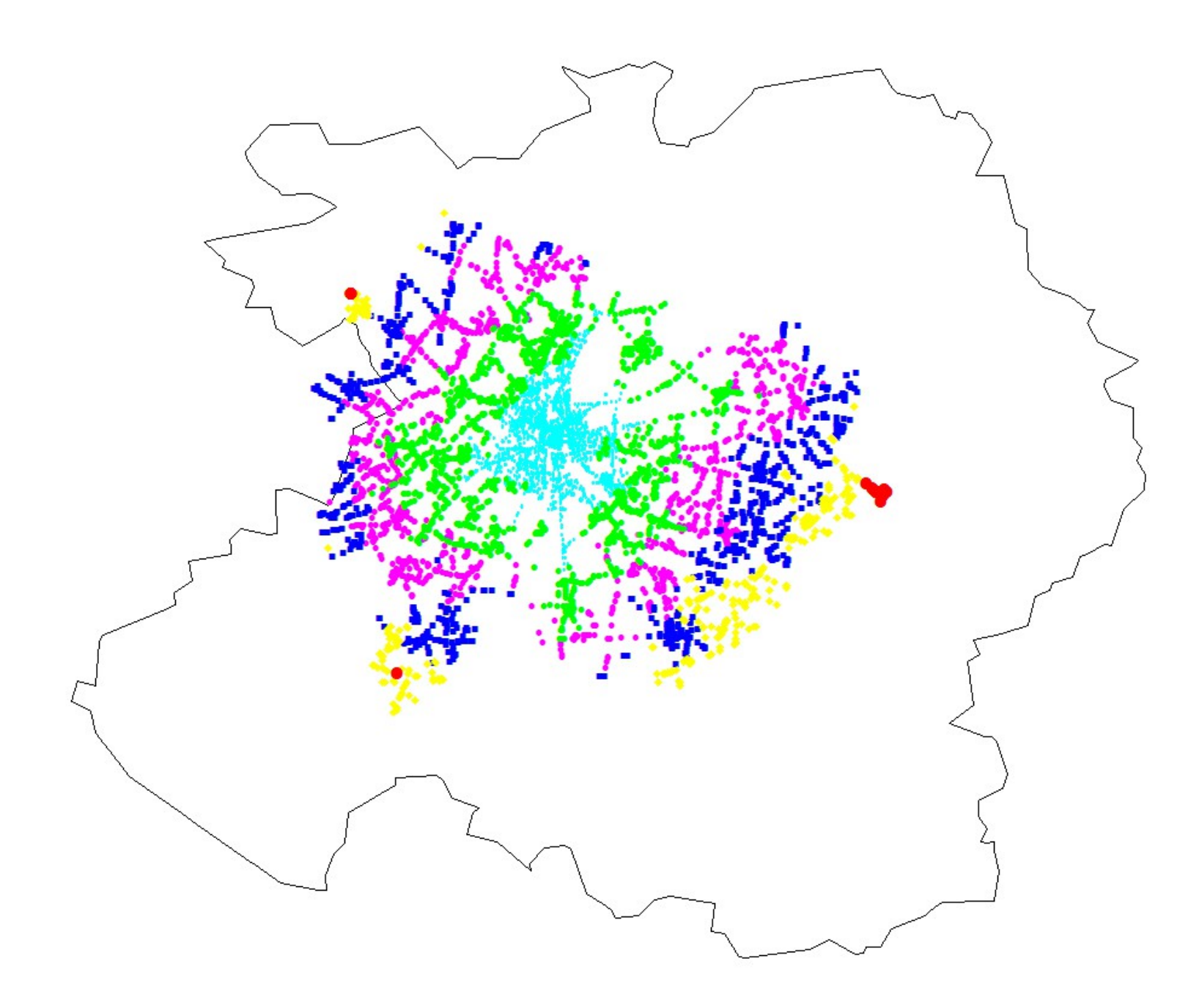}}  
  \subfloat[Risk zone]{\label{fig:foret}\includegraphics[width=0.33\textwidth]{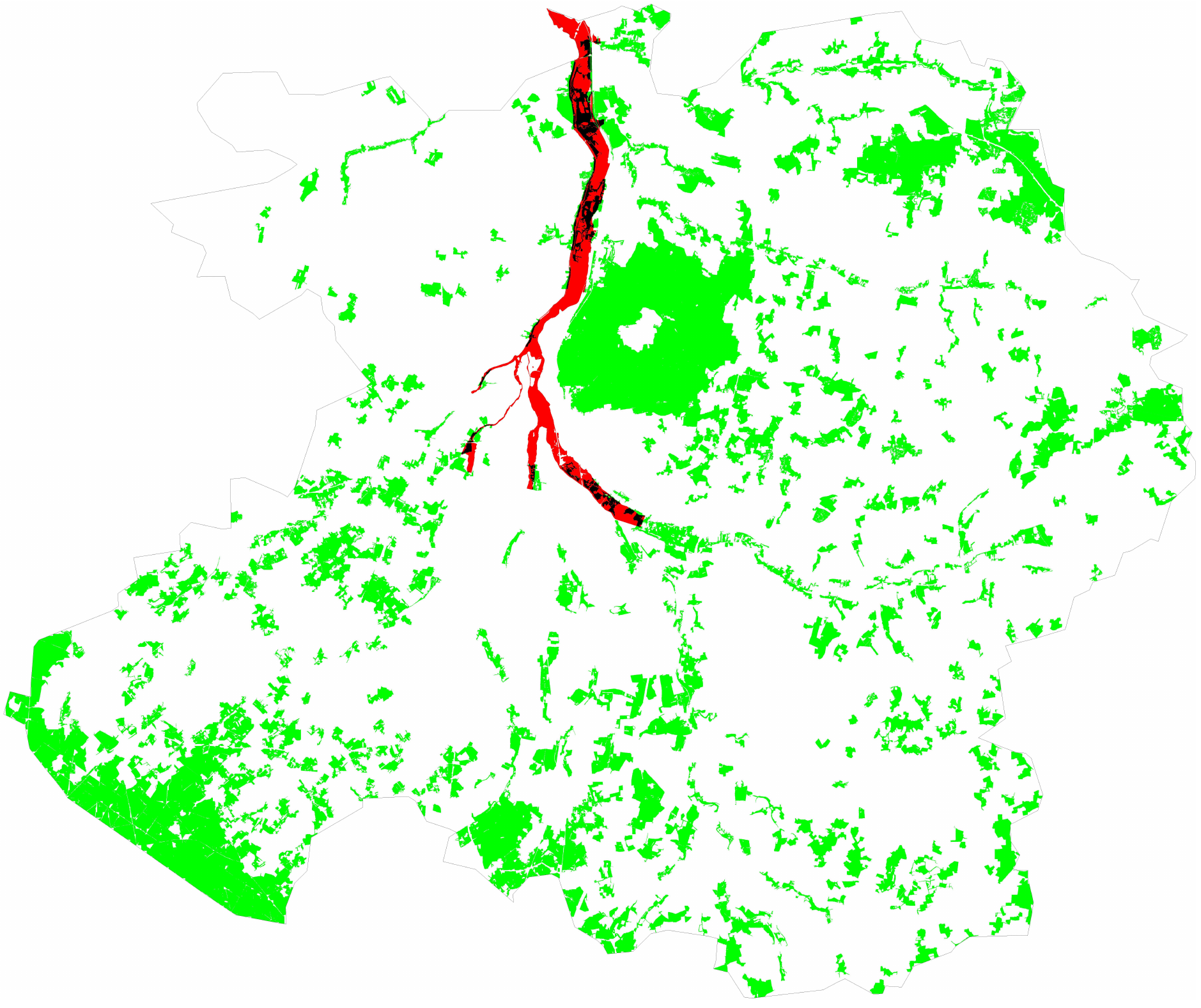}} 
  \caption{}
  
  \label{fig:parametre}
\end{figure}

A land lot (or a plot) is a land in the register of ownership (``cadastre'') with a determined area and boundary, owned by an individual or a company (see figure \ref{fig:parcelles}). Hence, land lots form a partition of the landscape. Their size and forms are variable, from a few square meters to more than $10^4 \; m^2$.

A plot that contains buildings will be called  occupied and we will use the term empty (or vacant) otherwise.

The plots are of different size and forms (we are no longer in the square lattice paradigm). The big plots which are usually far from the center of city can be divided into the smaller ones. However, due to the cost of division, a big plot is often divided all at once to be sold or constructed. In fact, we may admit that a big plot will be divided when it has a high construction probability (real estate value). We make the hypothesis that an occupied land lot will maintain its status in the future (it does not become vacant). These plots are the objects of our study. They replace sites (pixels) in the previous model.\\

This model does not consider size, shapes or type of buildings in land lots.
Since the type of the buildings  (commercial, administration, schools, hospitals ...) are not differentiated in the model,  their effect in city growth is assumed to be the same. It would not be difficult to take under consideration these different uses  but this would need a precise knowledge of them, which was not available to us.\\

Hence we will adapt the CGP model on land lots with the following assumptions:
\renewcommand\labelitemi{$\bullet$}
\begin{itemize}
  \item The change of status of a land lot (vacant to occupied) occurs with a probability related not only to distance to the center of the city but also to local density of the urban system and accessibility,
  \item An occupied land lot will stay occupied, there is no change from occupied to vacant,
  \item We do not distinguish the different types of buildings in land lots neither from size nor from purpose of use.\\
\end{itemize}

Let us come now to the details of the model. We divide the area of Montargis into a set $\mathcal{S}$ of $M\times N$ small squares  and we use them to pixelize the land-lots. From now on, all pixels on a given land-lot will be either all white or all black. We define the function ${\mathcal M}$ on the grid by
\[
\mathcal{M}\left(z\right)=\begin{dcases*}
        1  & if $z$ is part of an occupied land lot,\\
        0 & eitherwise.
        \end{dcases*}
\]
Let $\phi(t)=e^{-|t|}$, $c=\displaystyle\sum_{z\in{\mathcal S}}\phi(z)$ and $\varphi=\frac1c\phi$.
We define the local density as 
\begin{equation}\label{Delta}
\Delta(z)={\mathcal M}\star\varphi(z)=\frac{1}{c}\sum_{{\mathcal M}(z_1)=1}e^{-|z-z_1|}.
\end{equation}

We also assume that the probability of construction of a land lot is related to its \textit{accessibility}, which we now define: first we define a set $\mathcal N$ of ``nodes'', i.e. points on the road network  (mainly cross-roads). Following \cite{Gane}, the accessibility  $A(z)$ of a given pixel $z$ is a (possibly ponderated) sum over all the nodes $z'$ of the road network of the time  $t(z,z')$ necessary to go to  $z'$  from the node closest to $z$ (figure \ref{fig:noeuds}):
\begin{equation}\label{alpha}
A(z)=\sum_{z'\in {\mathcal N}}t(z,z')
\end{equation}
As for density, locations with high accessibility (small $A$) will be considered as more attractive. 

Finally we make the assumption  that being close to a river and/or to the center of the city does attract settlements. 

All these considerations lead us to modify the probability  $p_t$ defined in (\ref{pi}) as follows:

		\begin{equation}
		p_t(z) = \frac{1}{C}e^{-\lambda_{t,1} |z-z_0|-\lambda_{t,3} d(z,R)}\;\Delta(z)\; e^{-\lambda_{t,2} A(z)}
		\label{eq:3ocprobf}
		\end{equation}
		where
		\begin{itemize} 
		\item $C$ is a normalizing constant,
		\item $\lambda_{t,1}$, $\lambda_{t,2}$, $\lambda_{t,3}$ are  parameters possibly depending on time $t$ that will be fixed empirically using the historical data of Montargis,
		\item $z_0=0$ is the center of the city,
		\item $d(z,R)$ is the shortest distance from $z$ to the river,
		\end{itemize}
	

To summarize, the CGP model is used on {\it vacant} land lots which become {\it occupied} with probability given by \eqref{eq:3ocprobf}. The correlation exponent in the model has been empirically chosen to be  small  ($\alpha = 0.001$, large correlations). 
The starting point  $\mathcal{M}_0$ is Montargis in 1900. The simulation builds  ${\mathcal M}_{i+1}$ from ${\mathcal M}_i$ only: no external information is added.\\

\subsection{Simulations}
\begin{figure}[h!]
	\centering
		\includegraphics[width=0.5\textwidth]{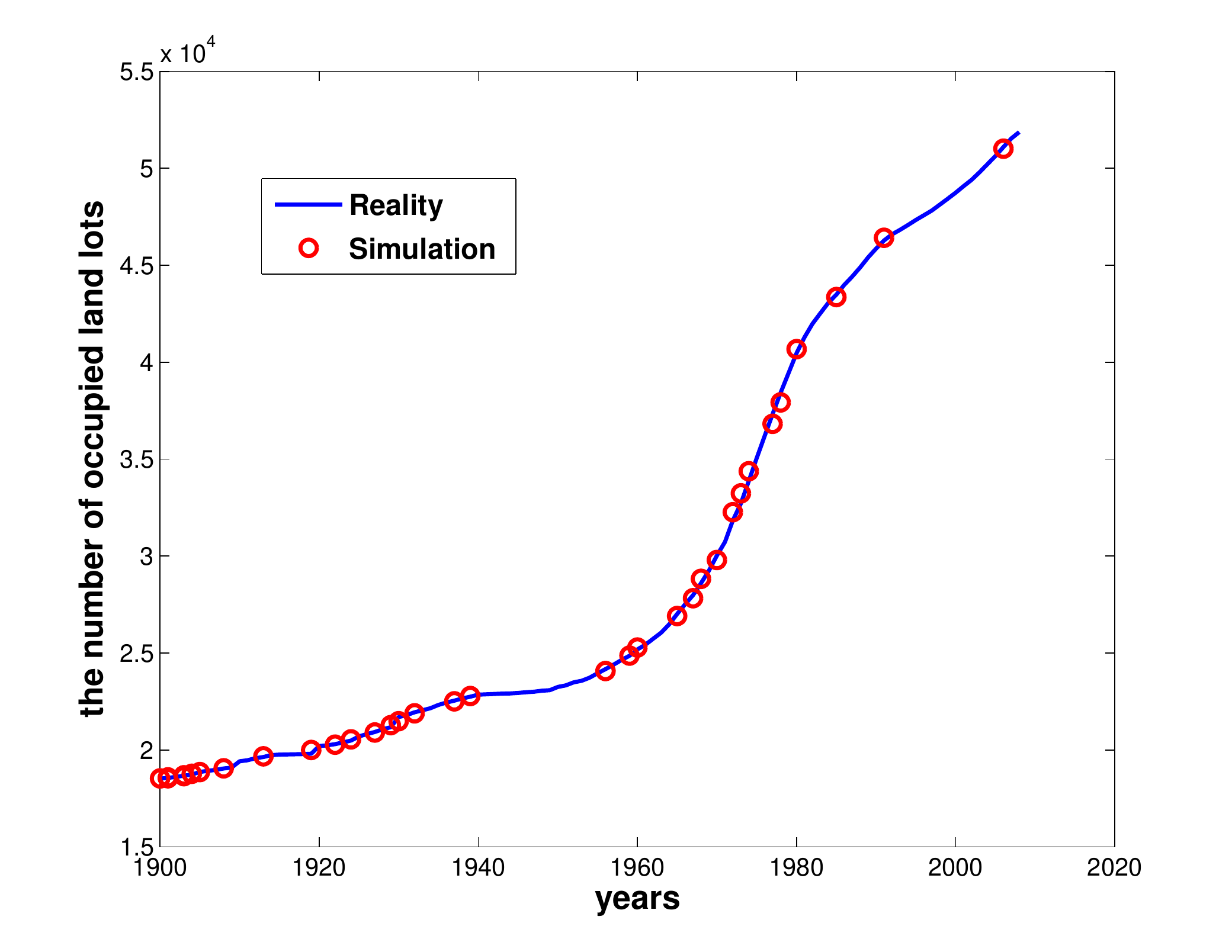}	
		\caption{The evolution of the number of constructed plots of real Montargis and its simulations from 1900 to 2008.}
		\label{fig:3evoluyears}
\end{figure}

The model cannot predict the rythm of the city growth, which is clearly non-linear: for instance the war period is very different from the sixties' which is itself very different from the beginning of the XXth century. Thus we change time in such a way that the total number of building coincides between reality and simulations at each time (figure \ref{fig:3evoluyears}).

\begin{figure}[h!]
  \centering
  \subfloat[Montargis 1949]{\label{fig:mon1949}\includegraphics[width=0.4\textwidth]{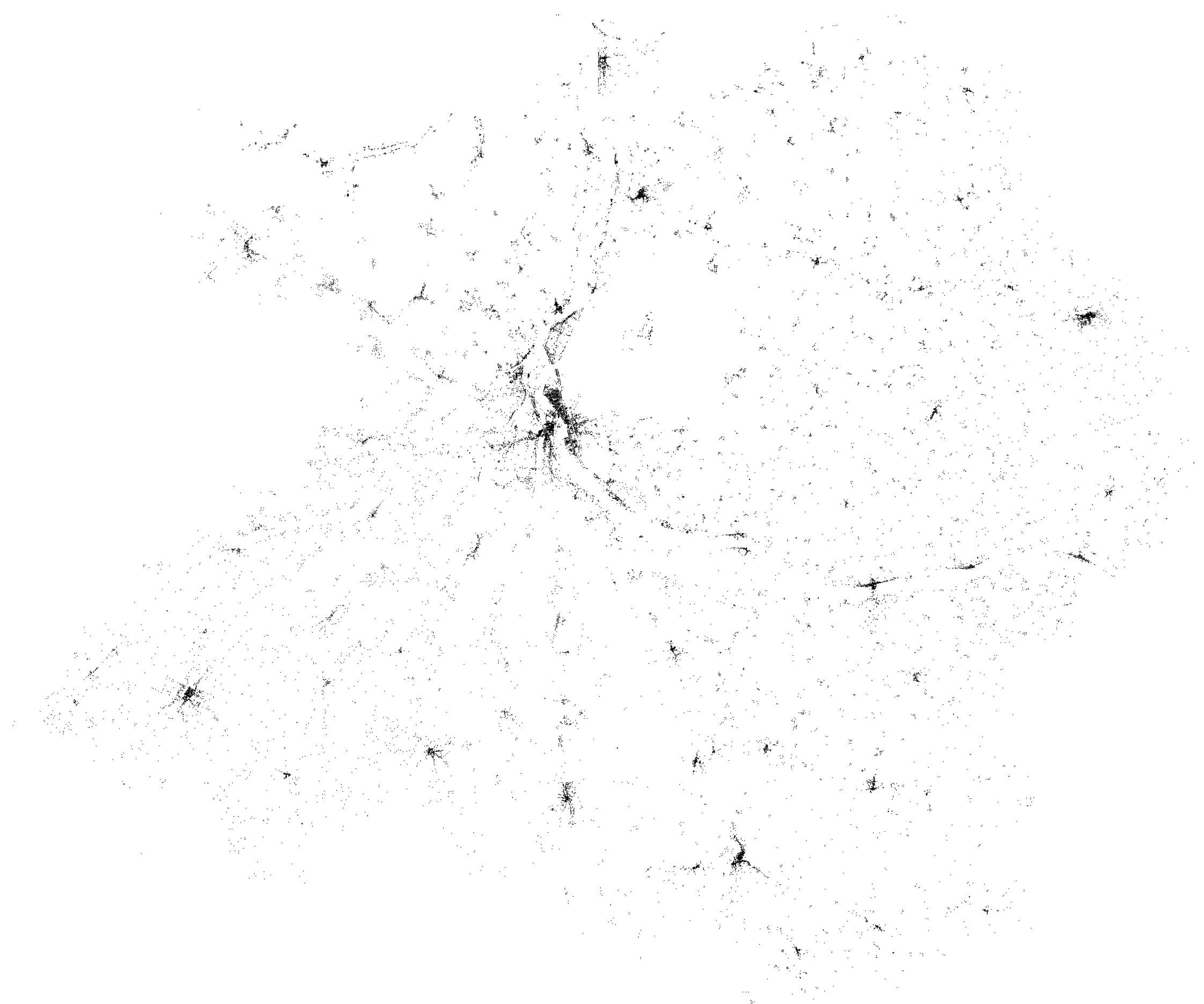}}
  \subfloat[Simulation]{\label{fig:simMon22}\includegraphics[width=0.4\textwidth]{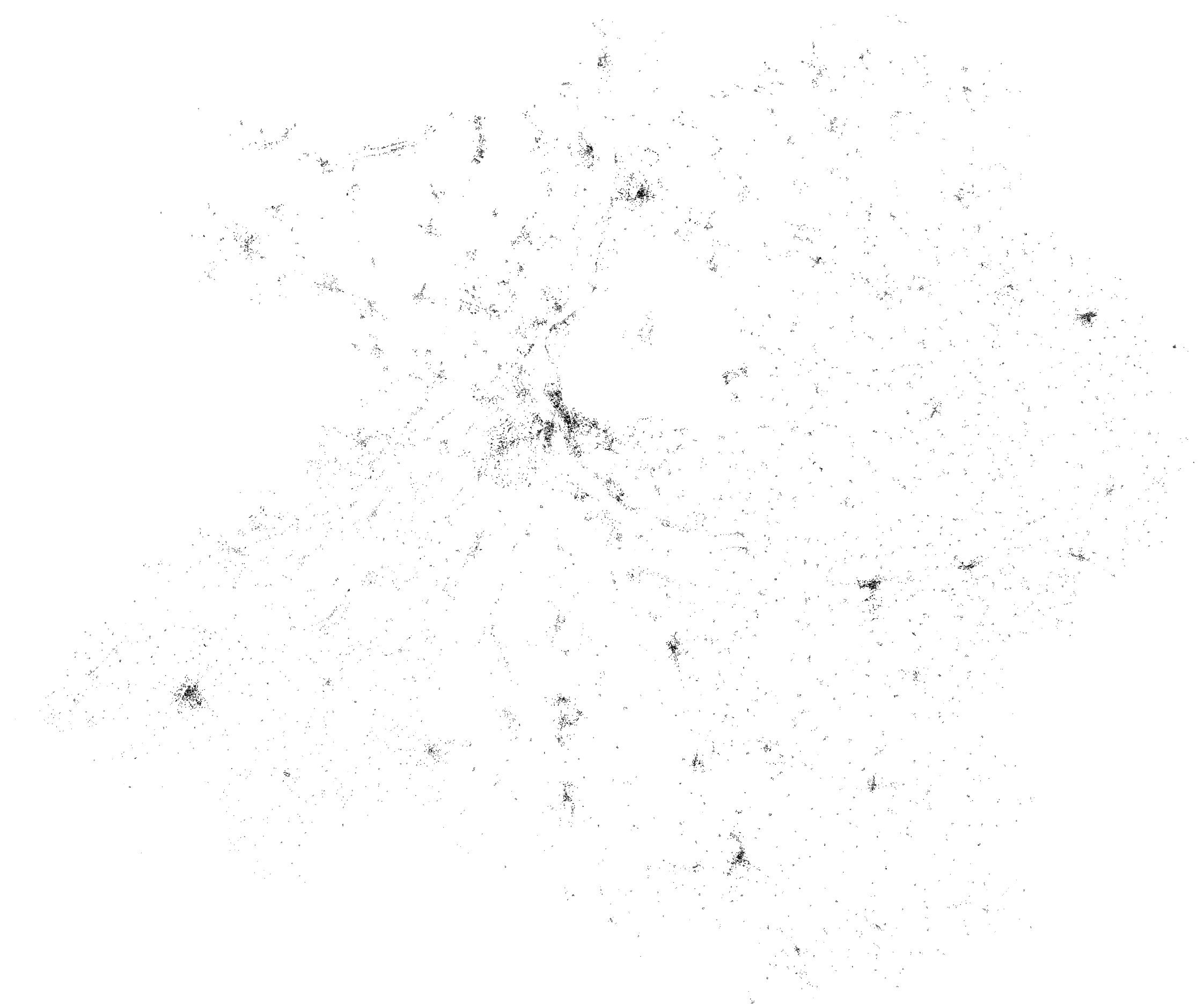}}\\
  \subfloat[Montargis 1979]{\label{fig:mon1979}\includegraphics[width=0.4\textwidth]{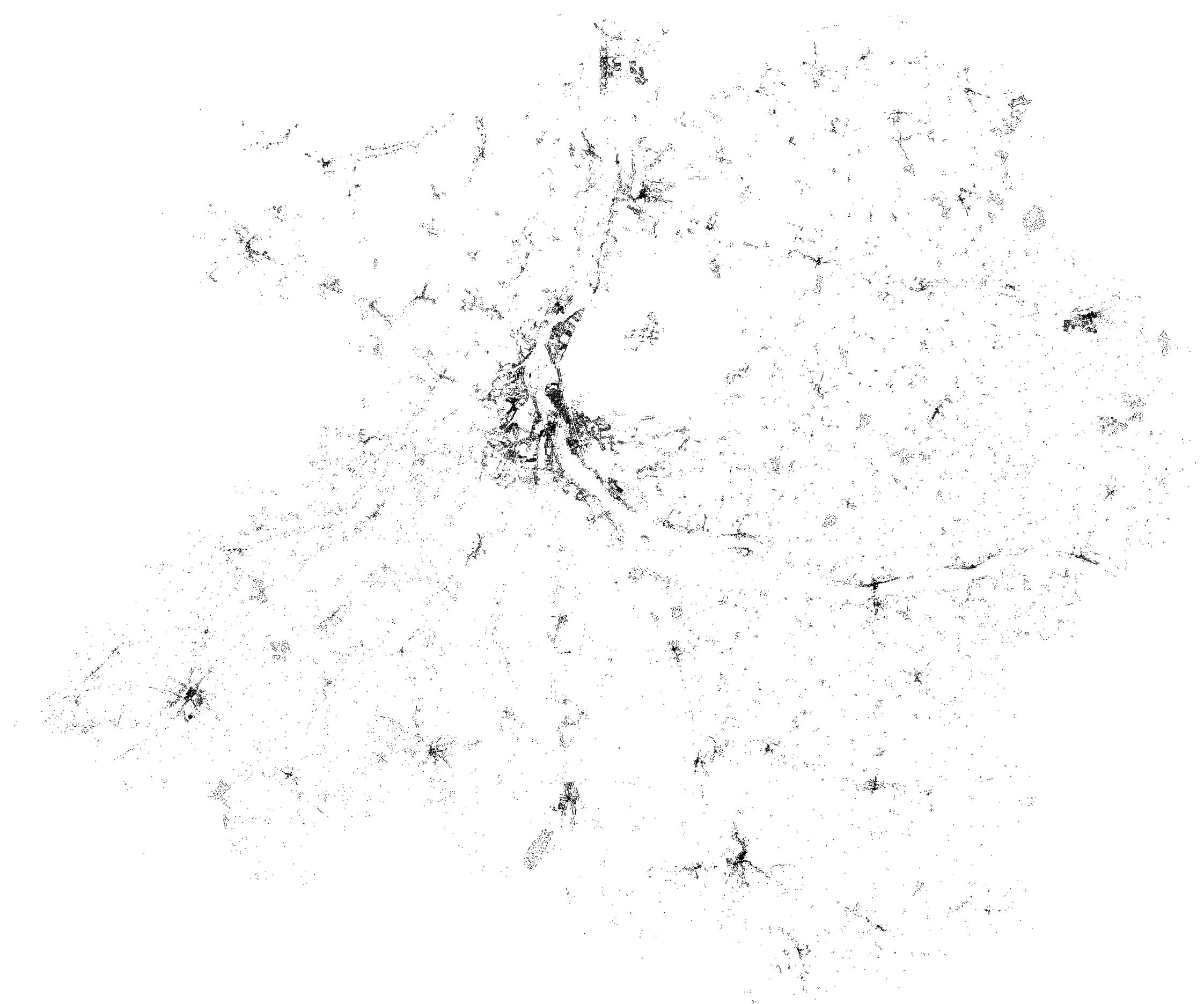}}
  \subfloat[Simulation]{\label{fig:simMon33}\includegraphics[width=0.4\textwidth]{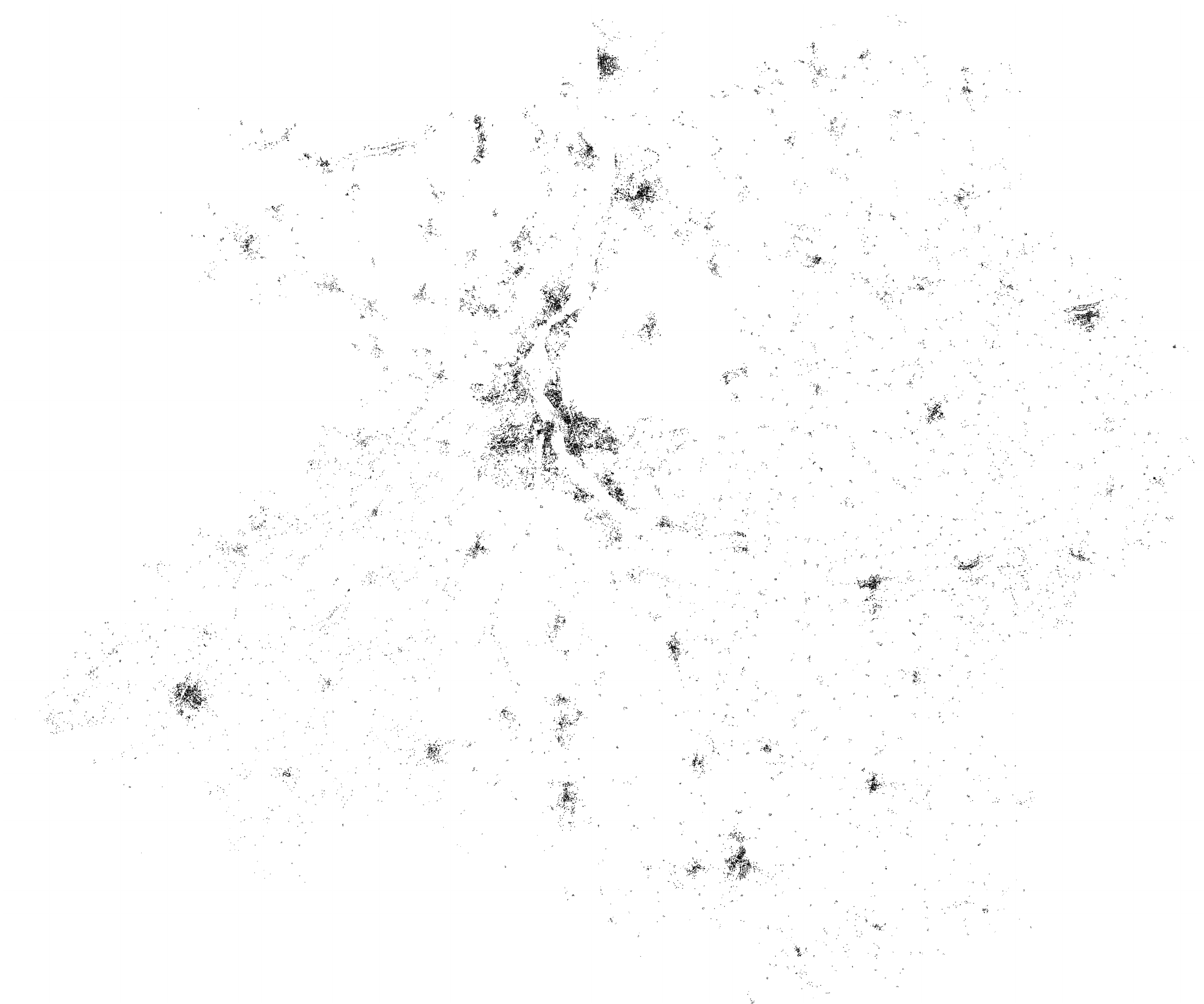}}\\
  \subfloat[Montargis 2007]{\label{fig:mon2007}\includegraphics[width=0.4\textwidth]{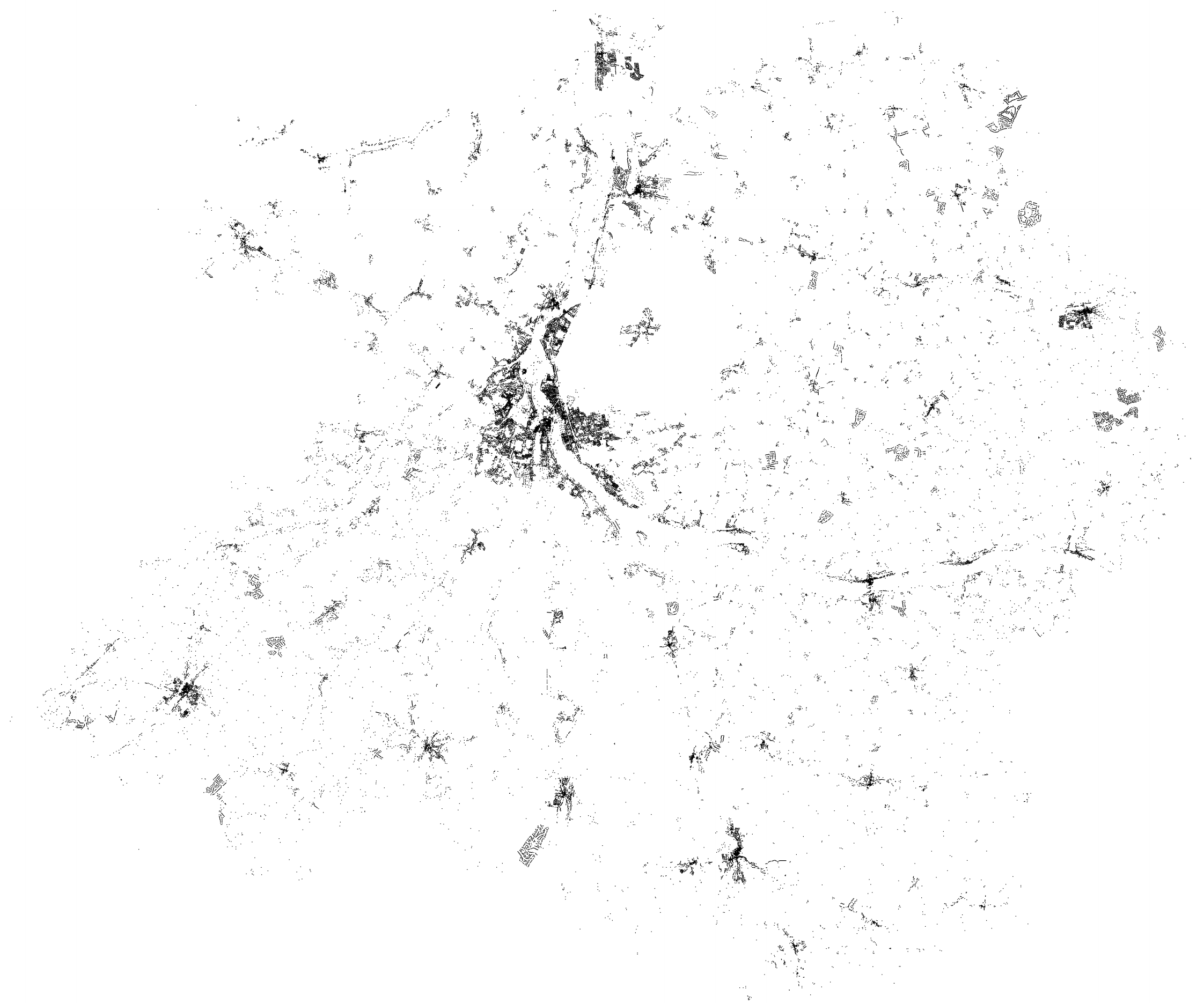}}
  \subfloat[Simulation]{\label{fig:simMon37}\includegraphics[width=0.4\textwidth]{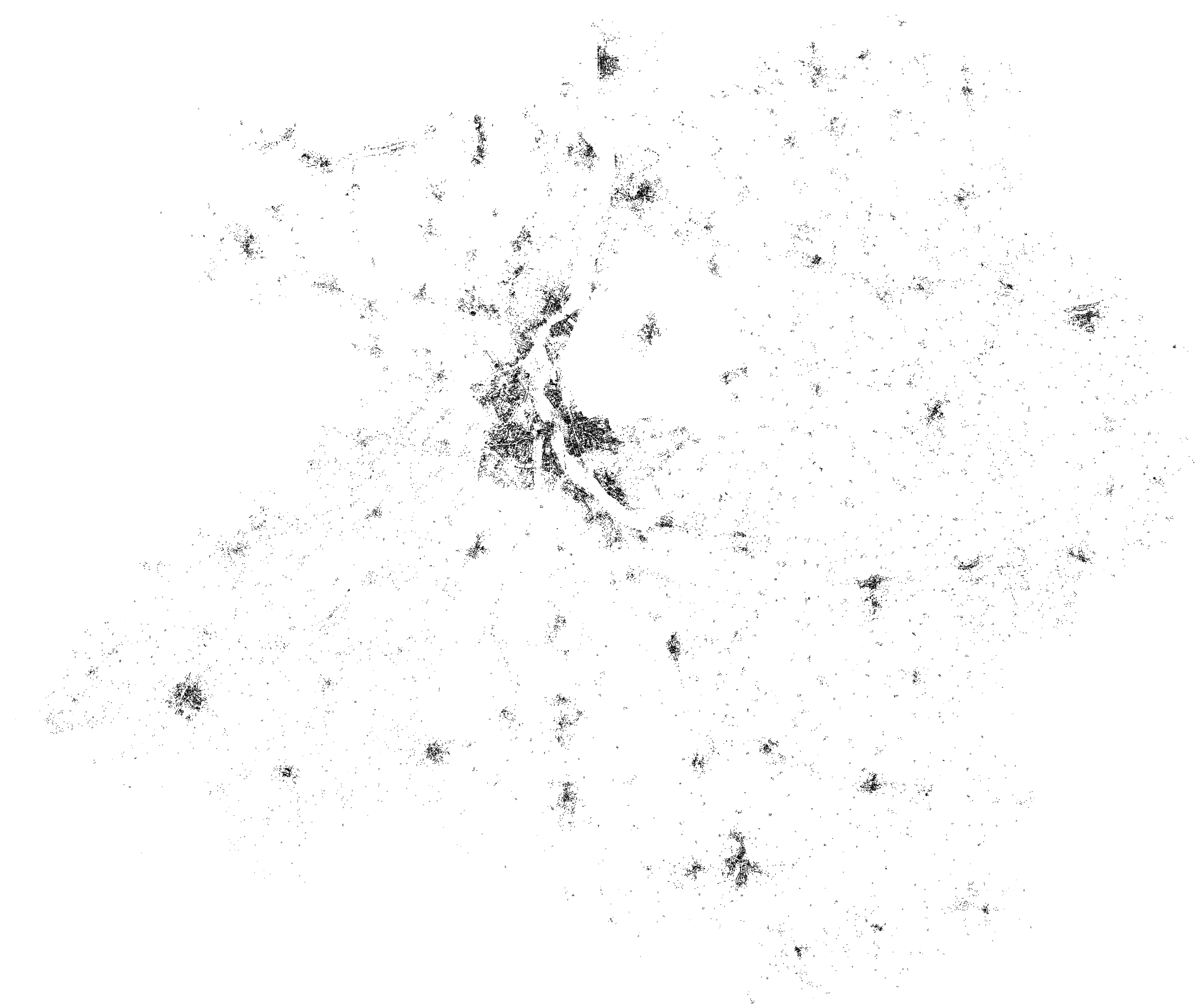}}
  \caption{Real Montargis for the years 1949, 1979, 2007 (on the left) and the corresponding simulations (on the right) realized by using Correlated Gradient Percolation with the occupation probability function \eqref{eq:3ocprobf}.}
  \label{fig:simMon}
\end{figure}

Starting from Montargis 1900 (figure~\ref{fig:Mon1900})  we obtain the simulations for the growth of Montargis from 1900 to 2007. In the figures \ref{fig:simMon}  we present the simulations at 3 equidistributed times. All the parameters present in the formula (\eqref{eq:3ocprobf}) for $p_t$ are empirically adjusted in order to best fit with reality. This choice is then applied once for all.

\subsection{Comparisons}
\begin{figure}[ht!]
	\centering
		  \subfloat[The overlap between real Montargis (red) and its 2008 simulation (blue)]{\label{fig:overlap}\includegraphics[width=0.5\textwidth]{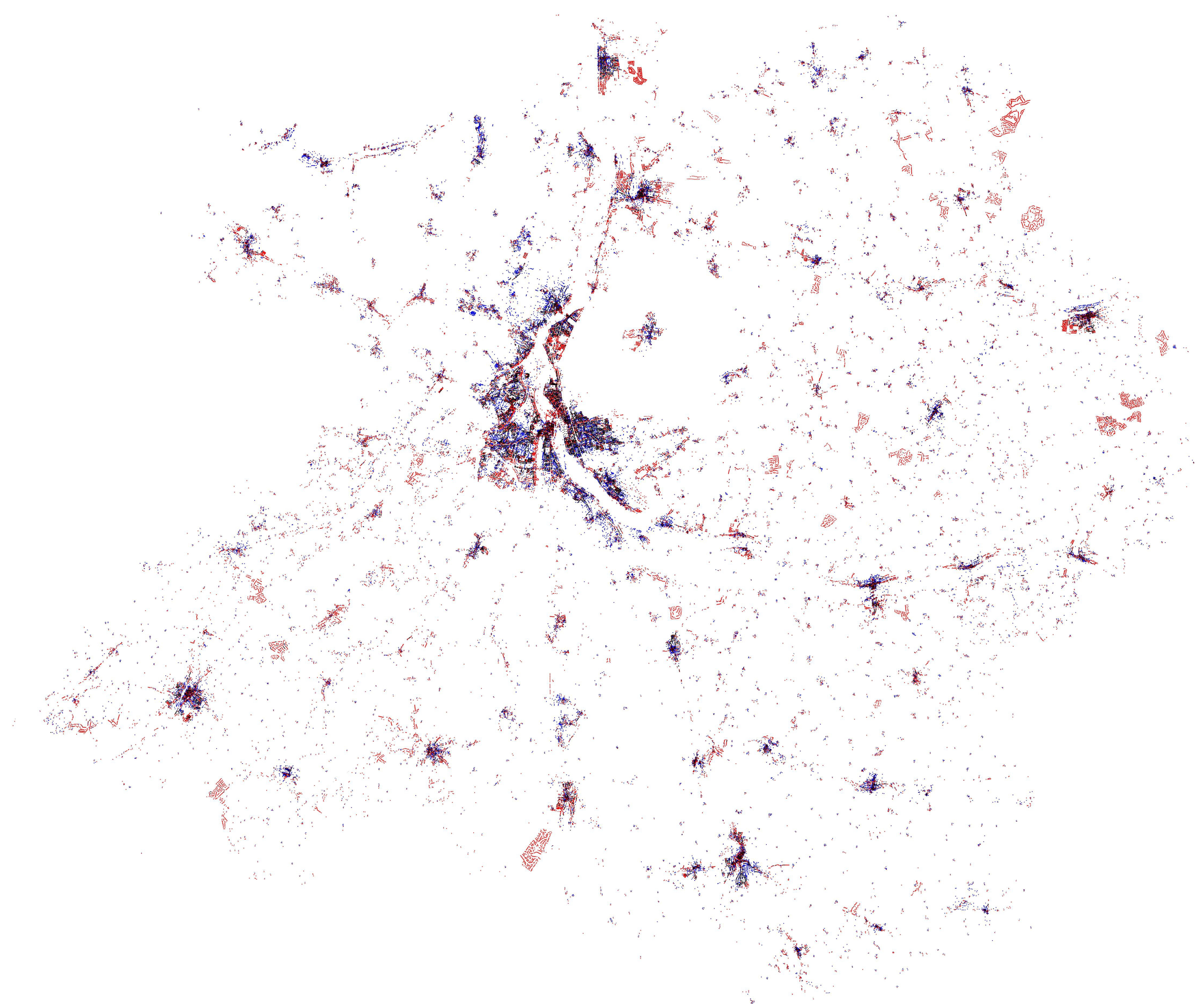}}
		 \subfloat[The distribution of errors]{\label{fig:errordist}\includegraphics[width=0.5\textwidth]{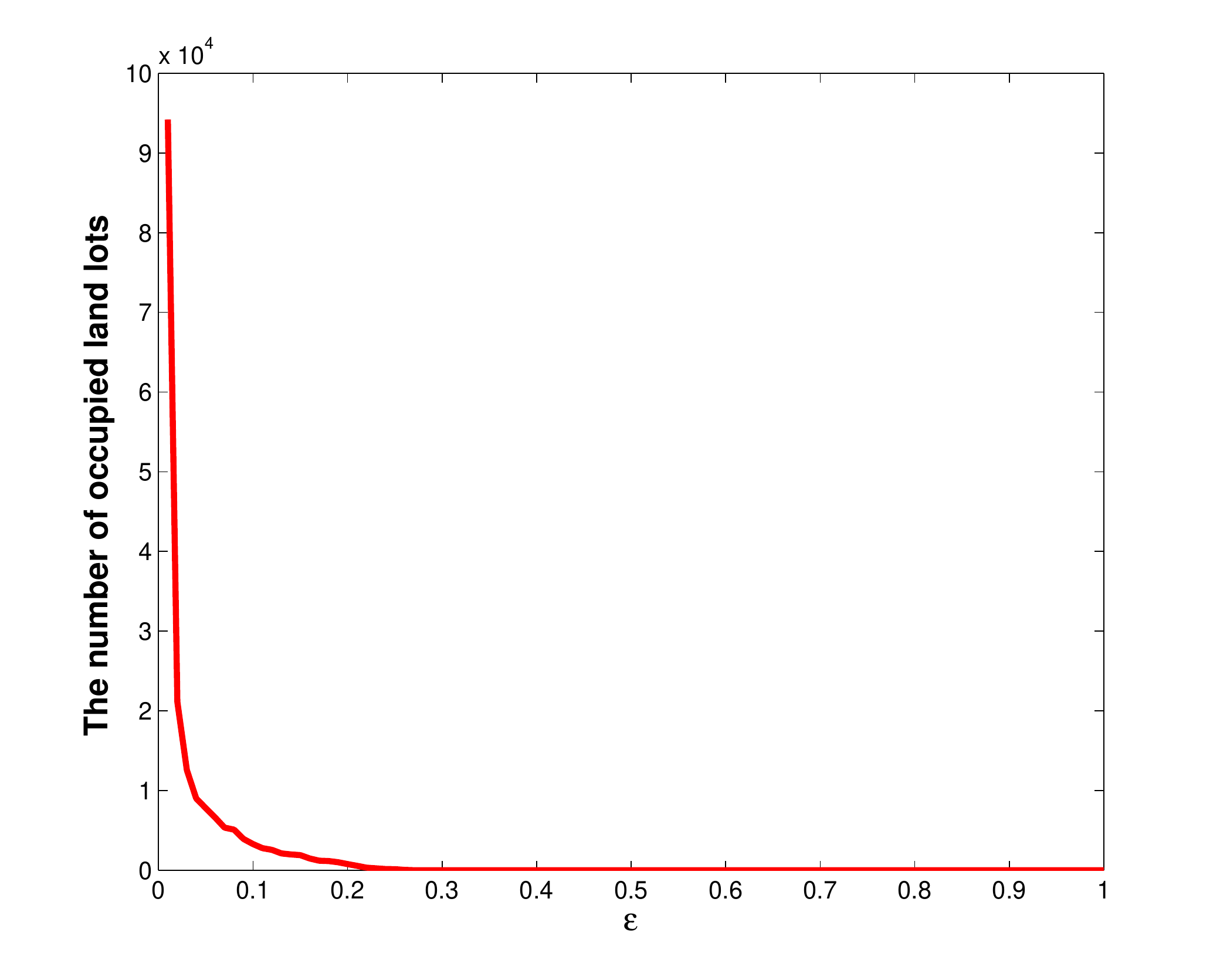}}
		 \caption{Quantitative comparisons}
\end{figure}

A comparison is performed by overlapping real Montargis 2008 with its simulation (figure \ref{fig:overlap}). There are some regions near the north boundary where the simulations do not match as well as near the center. In fact,  the speed of growth is higher there due to proximity of Paris and the emerging of peri-urban population.\\
Also, from place to place, there have been new subdivisions of  blocks of buildings that of course could not be predicted.
Besides these two exceptions, simulations fit rather well with reality, as quantified in the next section.



\subsection{Quantification of the error term in the simulations}
In addition to the above overlap, we give some numerical comparisons between Montargis 2007 and its simulation.\\
Let $\Delta_r(z)$, $\Delta_s(z)$ be the density functions of a plot $z$ in real Montargis ${\mathcal M}_r$ and its simulation  ${\mathcal M}_s$ respectively; i.e. 
\[
\Delta_j(z) = \frac1c\sum_{z'\in {{\mathcal M}_j}}e^{- |z-z'|}, \quad j=r,s
\]
We define the relative error at $z$ as 
$$\tau(z)=\frac{|\Delta_r(z)-\Delta_s(z)|}{\max\limits_z\left\{\Delta_r(z);\Delta_s(z)\right\}}$$

The figure \ref{fig:errordist} shows the distribution of these relative errors of density.

The maximal relative error of density is
\[
\epsilon_{max}=\max_{z}{\tau(z)}=\textcolor{red}{0.25},
\]
the mean error is
\[
\epsilon_{mean}=\frac{1}{n}\sum_{z}{|\tau(z)|}=\textcolor{red}{0.031}
\]
and the mean squared error is
\[
\epsilon_{MSE}=\frac{1}{n}\sum_{z}{|\tau(z)|^2}=\textcolor{red}{0.003}.
\]
where $n$ is the  number of terms in the sum.

\section{Conclusion}

Using the information that has been made available to us by the local authority, we have been able to construct a powerful and realistic model for the city growth of Montargis that could serve as a useful tool for decisions makers. This multi-parametrized model is also very flexible and can thus be applied to other cities as well. Compared with the inspiring model of \cite{H.Makse1998}, it introduces new quantified notions such as density or accessibility which have proven to be pertinent in connection with city growth, at least for middle sized French cities. It would be interesting to investigate to what extent it might be applied to the growth of modern megapoles which seem to obey different rules whose quantification is a challenge: we intend to explore, in the near future, the case of overgrowing African megapoles that look set to change African landscape and culture.\\
Another continuation of this model would be to apply it at a different scale, such as networked cities: a first attempt has been done for the French \textit{R\'egion Centre}, presented at  the centennial conference of the American Geophysics Union (\cite{BNZ}).
Besides its possible further uses, this model also raises many theoretical questions. As we have seen, among  the main quantities that rule the models are local density and the correlation coefficient $\alpha$. These two quantities are pretty much related: In the simulations it appears that the correlation factor seems to favor the emergence of "seeds" of Christaller satellites while the density factor transforms these seeds in real "sub-towns". It would be interesting to understand further this phenomenon and the mutual influence of these factors. \\
In a work in progress we investigate how one could statistically estimate the correlation coefficient $\alpha$. It is in particular unclear if this can be done with the sole knowledge of the city map at a given time or if we really need the whole history of the town.

\bibliographystyle{apsr}
\bibliography{BNZ}

\end{document}